\documentclass[twocolumn,aps,prb,superscriptaddress,showpacs]{revtex4}

\usepackage{amsmath,amssymb}
\usepackage{graphicx}

\newcommand\F{\scriptscriptstyle \rm F}

\begin{document}

\title{Wave-packet Formalism of Full Counting Statistics}

\author{F. Hassler}
\affiliation{Theoretische Physik,
ETH Zurich, CH-8093 Zurich, Switzerland}

\author{M.V. Suslov}
\affiliation{Institute of Solid State Physics RAS, 142432 Chernogolovka,
Moscow Region, Russia}

\author{G.M. Graf}
\affiliation{Theoretische Physik,
ETH Zurich, CH-8093 Zurich, Switzerland}

\author{M.V. Lebedev}
\affiliation{Institute of Solid State Physics RAS, 142432 Chernogolovka,
Moscow Region, Russia}

\author{G.B. Lesovik}
\affiliation{L.D.\ Landau Institute for Theoretical Physics RAS,
117940 Moscow, Russia}

\author{G. Blatter}
\affiliation{Theoretische Physik,
ETH Zurich, CH-8093 Zurich, Switzerland}

\date{\today}

\begin{abstract}
  We make use of the first-quantized wave-packet formulation of the full
  counting statistics to describe charge transport of noninteracting
  electrons in a mesoscopic device. We derive various expressions for
  the characteristic function generating the full counting statistics,
  accounting for both energy and time dependence in the scattering process
  and including exchange effects due to finite overlap of the incoming
  wave packets. We apply our results to describe the generic statistical
  properties of a two-fermion scattering event and find, among other
  features, sub-binomial statistics for nonentangled incoming states
  (Slater rank 1), while entangled states (Slater rank 2) may generate
  super-binomial (and even super-Poissonian) noise, a feature that can be
  used as a spin singlet-triplet detector. Another application is concerned
  with the constant-voltage case, where we generalize the original result
  of Levitov-Lesovik to account for energy-dependent scattering and finite
  measurement time, including short time measurements, where Pauli blocking
  becomes important.
\end{abstract}


\maketitle

\section{Introduction}

Charge transport across an obstacle in a wire is a statistical process,
whose complete description is provided by the probability function
$P(n,t)$, telling how many charge carriers $n$ are transmitted through
the wire during the time $t$. The calculation of this full counting
statistics usually aims at the generating function $\chi(\lambda,t)= \sum_n
P(n,t)\, e^{i \lambda n}$ for this process, from which the probability
distribution $P(n,t)$ follows through simple Fourier transformation
${\cal F}[\chi(\lambda,t)] = P(n,t)$. The proper physical definition of
the generating function $\chi(\lambda,t)$ is a nontrivial problem and
has been solved by Levitov and Lesovik back in 1993,\cite{levitov:93} see
also Ref.~\onlinecite{levitov:96}, with numerous applications to follow
\cite{nazarov}. The original definition includes a `charge counter' in the
form of a spin, coupled via the gauge potential to the moving charges, and
has been cast in a second-quantized formalism of appreciable complexity. The
recent observation \cite{lesovik:06} of the correspondence between the
generating function $\chi_1(\lambda)$ of the full counting statistics
for one particle and the notion of fidelity in a (one-particle, chaotic)
quantum system \cite{peres:84} has lead to a much simpler first-quantized
formulation of full counting statistics, including the generalization
$\chi_N(\lambda)$ to $N$ particles. In fact, a first-quantized version
of charge transport to calculate noise has been already introduced some
years ago. \cite{martin:92} Furthermore, such a wave-packet formalism
naturally describes the statistics of pulsed transport, where unit-flux
voltage pulses generate single-particle excitations feeding the device of
interest \cite{lee:95,levitov:96,ivanov:97,lebedev:05,keeling:06} (a source
injecting individual electrons into a quantum wire has been realized in a
recent experiment \cite{feve:07}). The simplicity of the first-quantized
formalism then has allowed to obtain nontrivial results on the full counting
statistics for an energy dependent scatterer, including its dependence on
the exchange symmetry of the transported charge. \cite{hassler:07}

In this paper, we make intense use of this wave-packet formalism of charge
transport and (re-)derive various expressions for the characteristic
function $\chi_N(\lambda)$ in a much simplified manner.  We start with an
$N$-particle Slater determinant made from orthonormalized single-particle
wave functions $\phi_m$ describing fermions incident from the left and
derive the associated characteristic function describing the full counting
statistics in determinant form,
\begin{equation}\label{eq:fcs_0}
  \chi_N(\lambda) = \det\langle \phi_m|1-\mathcal{T}+
  \mathcal{T}e^{i\lambda}|\phi_n\rangle,
\end{equation}
with the operator $\mathcal{T}$ describing the energy dependent transmission
across the scatterer, $\mathcal{T} = \int (dk/2\pi) T_k |k\rangle\langle k|$
in momentum ($k$) representation (here, the particle number $N$ replaces
the time variable $t$ in the original formula \cite{levitov:93}). The
determinant in Eq.~(\ref{eq:fcs_0}) can be cast in a product form
\begin{equation}\label{eq:fcs_1}
  \chi_N(\lambda) = \prod_{m=1}^N (1 - \tau_m + \tau_m e^{i\lambda}),
\end{equation}
where $\tau_m$ are the eigenvalues of the Hermitian operator $\mathcal{T}$
in the space spanned by the basis states $|\phi_n\rangle$. We denote the
distribution in (\ref{eq:fcs_1}) as \textit{`generalized binomial'}.

In a real experiment, the unit-flux voltage pulses generating the incoming
wave packets may overlap. For this situation, we rederive the simple and
elegant expression (\ref{eq:fcs_1}) for the full counting statistics, but
with the coefficients $\tau_m$ now replaced by the roots of a generalized
eigenvalue problem incorporating all effects of fermionic statistics and
the full energy dependence of the transmission. The results (\ref{eq:fcs_0})
and (\ref{eq:fcs_1}) apply to a nonentangled incident state in the form of
a Slater determinant; \cite{entangled} an extension to include entangled
states of Slater rank 2 is provided as well.  \cite{taddei:02}

Next, we generalize the result (\ref{eq:fcs_0}) to describe a setup where
both the scattering process and the counting window depend on time and
find a compact result in the form (\ref{eq:fcs_0}) with
\begin{equation}\label{eq:fcs_2}
  \mathcal{T} \rightarrow \mathcal{T}_Q = \mathcal{U}^\dagger 
  \mathcal{Q} \mathcal{U},
\end{equation}
where $\mathcal{U}$ denotes the single-particle time evolution operator
and the operator $\mathcal{Q}$ projects the wave function onto its measured
(counted) part.  Full counting statistics for fermionic atoms in determinant
form has been derived in Ref.~\onlinecite{glauber} through transcription
of the bosonic expression \cite{glauber:65} to the fermionic case, see
Ref.~\onlinecite{braungardt:08} for a recent application.

Finally, we extend the result Eq.~(\ref{eq:fcs_2}) to the situation where
the incoming state consists of an incoherent superposition of many Slater
determinants with different particle numbers. For the case of particles
incident only from the left side, we find the result (\ref{eq:fcs_0}) with
\begin{align}\label{eq:fcs_3}
  \mathcal{T} \rightarrow \eta \mathcal{T}_Q,
\end{align}
where $\eta$ denotes the one-particle occupation-number operator. In
addition, the determinant in (\ref{eq:fcs_0}) has to be taken over all
the single-particle Hilbert space.

We make extended use of these formulas: for a two-particle problem, we show
that, {\it i)} an incoming state described via a simple Slater determinant
cannot generate a Fano factor $F = \langle\langle n^2 \rangle\rangle/
\langle n \rangle > 1-\langle n\rangle/2$ (i.e., noise is always sub-binomial
and in particular also sub-Poissonian; there is no bunching); the above
cumulants are obtained through the generating function $\chi (\lambda)$
via $ \langle\langle n^j \rangle \rangle = (-i)^j \partial_\lambda^j\log
\chi |_{\lambda=0}$, {\it ii)} upon proper choice of $T_k$, an entangled
incoming state can generate any value for the Fano factor $F < 2$, and
{\it iii)} for two spin-$1/2$ fermions, we show that a simple scattering
experiment provides information on the entanglement of the incoming state
(cf.\ also Ref.\ \onlinecite{burkard:00}).

Subsequently, we analyze the situation with $N$ fermions and derive the
full counting statistics for a constant voltage ($V$) drive, thereby
generalizing the original result of Levitov and Lesovik \cite{levitov:93}
to describe transport with an energy dependent scattering transmission
(cf.\ Ref.\ \onlinecite{schonhammer:07}). Our result,
\begin{equation}\label{eq:fcs_4}
  \log \chi_{N} (\lambda) = N\frac{2\pi\hbar v_{\F}}{eV} \!
  \int_0^{e V/\hbar v_{\F}} \!\!  
  \frac{dk}{2\pi} \log(1- T_k + T_k e^{i \lambda}),
\end{equation}
admits the simple interpretation of the full counting statistics as deriving
from the transmission of the unbalanced Fermi sea residing between energies
$E_{\F}$ and $E_{\F}+eV$, with $E_{\F}$ denoting the Fermi energy and $V$
the applied bias. Using an alternative derivation based on (\ref{eq:fcs_2})
and stationary scattering states, we determine the short-time limit of
the counting statistics and rederive the binomial result (\ref{eq:fcs_4})
in the long-time situation, with the particle number $N$ replaced by the
measuring time $t$, $N \to t\, eV/2\pi\hbar$. The use of our determinant
formula combined with Szeg\H{o}'s theorem\cite{szego:15,szego:52} will
allow us to present a rigorous derivation of these results.

In the following, we give a short review of previous work on the
subject and then derive the characteristic functions (\ref{eq:fcs_0}) and
(\ref{eq:fcs_1}) of $N$ incoming fermions. In Sec.~\ref{sec:two_particles},
we apply these results to discuss the statistical transport-properties of
two fermions.  Section~\ref{sec:train} is devoted to the calculation of
the characteristic function for the constant voltage case starting from
$N$-particle trains and letting the width of the individual wave packets
go to infinity.  In Sec.~\ref{sec:generalizations}, we derive the results
(\ref{eq:fcs_2}) and (\ref{eq:fcs_3}) describing the setup involving a time
dependent scattering and counting incoherent superpositions of incoming
particles. We rederive the constant voltage result as an application,
including the short-time limit.

\section{Full Counting Statistics}\label{sec:fcs}

The first suggestion \cite{levitov:92} of a generating function for
full counting statistics relied on the straightforward expression
$\chi(\lambda,t) = \langle \exp [i \lambda \int dt'\, {\cal I}(t')]
\rangle$, where ${\cal I}(t)$ denotes the current operator. It then
was soon realized \cite{levitov:93} that this definition does not
correspond to any known (even on the level of a `Gedanken Experiment')
measuring procedure; still, this first definition produced the correct
results for all irreducible zero-frequency current-current correlators
$\langle\langle {\cal I}_0 \dots {\cal I}_0) \rangle\rangle$ (see also
the discussion in Ref.\ \onlinecite{lesovik:03}).  The first `practical'
definition \cite{levitov:94} of a generating function $\chi(\lambda,t)$,
corresponding (at least in principle) to a realistic counting experiment,
involved a spin-galvanometer as a measurement device (see also Ref.\
\onlinecite{levitov:96}). Recently, it has been pointed out \cite{lesovik:06}
that this suggestion (corresponding rather to a `Gedanken Experiment')
could actually be realized with qubits serving as a measuring device,
whereby the `environmental noise' generated by the transmitted charge
serves as the measurement signal for the full counting statistics. This
contrasts with the usual interpretation of the `environmental noise'
as being responsible for the qubit's dephasing \cite{neder:07} expressed
through the fidelity, and also relates to the competition between the gain
of information and dephasing \cite{averin:05} in quantum measurement theory.

The insight on the equivalence between the notions of fidelity and full
counting statistics has motivated a first-quantized formalism of the
counting problem in terms of wave packets. Fidelity $|\chi_\text{fid}|$,
the modulus of the overlap $\chi_\text{fid}=\langle \Psi_2 | \Psi_1
\rangle$, was introduces by Peres \cite{peres:84} in the context of
chaotic systems. It measures the overlap between two wave functions
$|\Psi_{1,2}\rangle$ which describe an initial state $|\Psi_0 \rangle$
which has evolved under the action of two slightly different Hamiltonians.
In the context of full counting statistics of a single particle measured by
a spin counter, the wave functions $\Psi_{1}$ and $\Psi_{2}$ are substituted
by scattering states $\Psi^{+}_\mathrm{out}$ and $\Psi^{-}_\mathrm{out}$
interacting with the spin counter in the states $|{\uparrow}\rangle$
and $|{\downarrow}\rangle$, resulting in an expression for the
generating function in the form $\chi_1 = \langle \Psi^{-}_\mathrm{out} |
\Psi^{+}_\mathrm{out} \rangle $.  This new first-quantized formulation in
terms of wave-packets provides a drastic simplification as compared to the
original second-quantized formalism.\cite{levitov:96} While the use of a
second-quantized formalism is mandatory for the description of particles
describing bosonic excitations of fields (photons, phonons, etc.), here,
we deal with nonrelativistic electrons where the particle number is fixed,
thus allowing for an alternative first-quantized description.  Moreover,
our wave-packet formalism has technical merits (e.g., in the description
of energy dependent scattering or in the classification of two-particle
scattering events) and also provides a better physical understanding. We
remark, however, that in dealing with finite temperatures we make use of
the second-quantized formalism in Fock space.

An alternative method, to the procedure based on a spin counter, was pursued
in several contributions \cite{muzykantskii:03,shelankov:03,avron:08} where
the full counting statistics and, in particular, its generating function
$\chi(\lambda,t)$, was constructed using only basic quantum mechanical
definitions; starting with an initial state in the form of an eigenstate
of the particle number operator with a fixed particle number to the right
of the scatterer (or the `counter'), a second projection (to eigenstates
of the number operator) onto the final state is carried out after the
observation time $t$.  Both procedures, projection and spin-counting, lead
to the same expressions for the generating function $\chi$, provided that the
incoming state involves no superposition across the scatterer. In the latter
situation, the explicit calculation using a spin-counter produces a fidelity
describing the decoherence of the spin, while an interpretation in terms
of a generating function can produce probabilities for noninteger charge
transport \cite{shelankov:03} and hence is unphysical.  On the other hand,
the projection method, destroying such a superposition in the course of the
first measurement, always admits an interpretation in terms of probabilities.

\subsection{One particle}

\begin{figure}[t]
  \centering
  \includegraphics{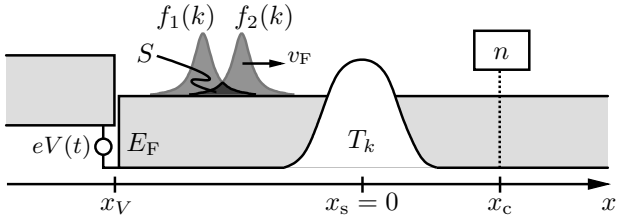}
  \caption{%
	Quantum wire with scattering center located at $x_\text{s}$
	giving rise to a momentum-dependent scattering probability $T_k$. A
	time-dependent potential $eV(t)$ applied at $x_V$ (to the left of the
	scatterer) generates incoming wave packets $f_1$, $f_2$ with overlap
	$S=\langle f_2 | f_1 \rangle$. A counter, placed at $x_\text{c}$ (to
	the right of the scatter), measures the statistics of the number $n$
	of transmitted particles. In our analysis, we consider incoming wave
  packets with momenta $k > 0$ residing outside the Fermi sea. As a
  result, the Fermi sea, which is not accounted for in our analysis, is
  not disturbed in the asymptotic time limit. For finite measuring times,
  the presence of the Fermi sea generates additional equilibrium noise
  which we do not consider in this article.
  }
  \label{fig:setup}
\end{figure}

In this paper, we make extensive use of the first-quantized formulation of
the generating function: starting with a simple one-particle problem, we
exploit the equivalence between the notion of fidelity and full counting
statistics. \cite{lesovik:06} Consider an incoming wave packet $\psi(x;
t\to -\infty)$ from the left of the form
\begin{equation}\label{eq:wf_in1}
  \psi(x; t) = \int \frac{dk}{2\pi}\, \phi_1(k) e^{i k x - i \epsilon(k) t}
\end{equation}
with normalization $\int (dk/2\pi) |\phi_1(k)|^2 =1$, cf.\
Fig.~\ref{fig:setup}. In the following, we assume (for simplicity) a linear
spectrum $\epsilon = v_{\F} k$ with $v_{\F}$ the Fermi velocity; at low
temperatures and voltages the interesting physics usually takes place near
the Fermi points. The momentum $\hbar k$ and the energy $\hbar \epsilon$
are measured with respect to the Fermi momentum $k_{\F}$ and the Fermi
energy $E_{\F}$.  Here and below, the wave-packets include only momenta
with $k>0$ in order not to disturb the Fermi sea which is considered to
be the vacuum in our analysis.  The scatterer at $x=0$ is characterized
by momentum(energy)-dependent transmission (reflection) amplitudes $t_k$
($r_k$; particle reflection takes us to the branch $\epsilon = -v_{\F} k$,
with $k$ measured relative to $-k_{\F}$).  The spin- (or qubit-) counter,
placed to the right of the scatterer, contributes a phase-factor $e^{\pm
i \lambda/2}$ to the wave function, where the sign depends on the state
$|{\uparrow}\rangle$, $|{\downarrow}\rangle$ of the spin-counter. The
outgoing ($t\to\infty$) wave-function assumes the form (we place the
counter right behind the scatterer at $x=0$)
\begin{align}\label{eq:wf_out1}
  \psi_{\text{out}}^\pm (x; t) =& \int \frac{dk}{2\pi}  [r_k
  e^{-i k (x+ v_{\F} t)} \Theta(-x) \\
  &+ t_k e^{i k (x- v_{\F} t)} e^{\pm i \lambda/2} \Theta(x) ] \phi_1(k)
  \nonumber
\end{align}
and consists of reflected ($x<0$) and transmitted ($x>0$) parts; $\Theta(x)$
is the unit step-function. The fidelity $\chi_1(\lambda)$ is given by
the overlap of wave functions with slightly different perturbations in
their evolution, here, with coupling to opposite spin-configurations
$|{\uparrow}\rangle$ and $|{\downarrow}\rangle$,
\begin{align}\label{eq:chi1}
  \chi_1(\lambda) &= \int\! dx\, \psi_\text{out}^{-}(x;t)^* \,
  \psi_\text{out}^{+}(x;t) \nonumber\\
  &\stackrel{(t\to\infty)}{\longrightarrow} \int \frac{dk}{2\pi }\, (1 - T_k +
  T_k e^{i\lambda}) | \phi_1 (k)|^2 \nonumber \\ &= \langle\phi_1| 1-
  \mathcal{T} + \mathcal{T}e^{i\lambda}| \phi_1 \rangle;
\end{align}
in the asymptotic or long-time limit, the integration over space is
trivially done by exploiting the complete separation of the wave function
into transmitted and reflected parts. Furthermore, the time dependence
disappears as soon as the transmitted wave function has passed the counter.
The transmission probabilities $T_k = |t_k|^2$ are the eigenvalues of the
transmission operator $\mathcal{T}=\int (dk/2\pi) T_k |k\rangle \langle k|$.
Given the above specific coupling to a spin, the fidelity is equivalent
to the characteristic function
\begin{equation}\label{eq:fcs_def}
  \chi(\lambda) = \sum_m P_m e^{i\lambda m}
\end{equation}
of the full counting statistics as defined in Ref.\ \onlinecite{levitov:94},
where a spin-galvanometer has been used as a measuring device.
The Fourier-coefficients $P_m$ are the probabilities for transmitting
$m$ particles. For the simple example of one incoming particle only two
outcomes are possible, particle reflection with probability $P_0 = 1 -
\langle \mathcal{T}\rangle$ and particle transmission with $P_1 = \langle
\mathcal{T} \rangle$, where $\langle \mathcal{T}\rangle =\langle \phi_1 |
\mathcal{T} | \phi_1 \rangle$ denotes the average transmission probability.
Knowing the characteristic function, the cumulants $\langle\langle n^j
\rangle\rangle$ can be obtained as the coefficients in the Taylor series
of $\log \chi(\lambda)$,
\begin{equation}\label{eq:cum}
  \langle\langle n^j \rangle \rangle 
   = \Bigl(\frac{d}{i d \lambda}\Bigr)^j \log \chi(\lambda) \Big|_{\lambda=0}.
\end{equation}
The ratio $F= \langle\langle n^2 \rangle\rangle / \langle n \rangle$
between the second and the first cumulant, called Fano-factor, will be of
special interest later.

\subsection{$N$ particles}

Next, we extend the above description to $N$ particles with an incoming wave
function $\Psi(\mathbf{k})$ defined in momentum space; the vector $\mathbf{k}
= (k_1,\dots,k_N)$ specifies the $N$ momenta of the particles. We consider
independent particles without interaction which scatter independently.
After scattering, the outgoing wave function assumes the asymptotic
($t\to\infty$) form
\begin{align}\label{eq:wf_out}
  \psi_\text{out}^\pm &(\mathbf{x};t) = \Bigl\{ \prod_{m=1}^N
  \int \!\frac{dk_m}{2\pi}\, [r_{k_m}
  e^{-i k_m (x_m+ v_{\F} t)} \Theta(-x_m) \nonumber\\
  & \quad+ t_{k_m} e^{i k_m (x_m- v_{\F} t)} e^{\pm i \lambda/2} \Theta(x_m)
  ] \Bigr\}
  \Psi(\mathbf{k}),
\end{align}
i.e., the evolution is the product of the single-particle evolutions
in expression (\ref{eq:wf_out1}). The characteristic function of
the full counting statistics $\chi_N (\lambda) = \int \!d\mathbf{x}\,
\psi_\text{out}^-(\mathbf{x};t)^* \psi_\text{out}^+(\mathbf{x};t)$ then
can be cast into the form
\begin{equation}\label{eq:chiN}
  \chi_N (\lambda) = \Bigl\{ \prod_{m=1}^N \int \frac{dk_m}{2\pi}\, (1 -
  T_{k_m} + T_{k_m} e^{i\lambda}) \Bigr\} |\Psi(\mathbf{k})|^2.
\end{equation}

So far, we did not specify the specific type of incoming wave function. If
we limit ourselves to Slater determinant states composed of orthonormalized
single particle states $\phi_m$,
\begin{equation}\label{eq:slater_det}
  \Psi(k_1,\dots,k_N) = \frac{1}{\sqrt{N!}}\det \phi_m(k_n),
\end{equation}
the expression Eq.~(\ref{eq:chiN}) can be rewritten as a single determinant
(see Eq.~(\ref{eq:slater_to_det}))
\begin{align}\label{eq:chiN_det}
  \chi_N(\lambda) &= \det \int \frac{dk}{2\pi} \, \phi_m^*(k) (1-T_k + T_k
  e^{i\lambda}) \phi_n(k) \nonumber \\
  &= \det \langle \phi_m | 1 -\mathcal{T} 
   + \mathcal{T}e^{i\lambda} | \phi_n \rangle
\end{align}
involving the single-particle matrix elements $\langle\phi_m| \mathcal{O}|
\phi_n \rangle$ of the operator $\mathcal{O} = 1 -\mathcal{T} + \mathcal{T}
e^{i\lambda}$.

\subsection{Nonorthogonal basis}

In a physical realization of such a scattering experiment, one usually
does not populate orthogonal states as used in the above construction of
the Slater determinant.  E.g., in the setup of Fig.~\ref{fig:setup} the
electrons typically occupy states $f_1$ and $f_2$ with a finite overlap,
i.e., they are nonorthogonal.  Of course, an $N$-particle Slater determinant
can be constructed as well out of nonorthogonal states $|f_m\rangle$,
provided they are linearly independent, i.e., $\det \langle f_m |
f_n \rangle \neq 0$.  The properly antisymmetrized and normalized wave
function~(\ref{eq:slater_det}) then acquires the form
\begin{equation}\label{eq:slater_det_non}
  \Psi^f(k_1,\dots,k_N) = \frac{1}{\sqrt{N! \det \langle f_m | f_n \rangle}}
  \det f_m(k_n).
\end{equation}
Inserting this expression into (\ref{eq:chiN}) and repeating the calculation
that led to (\ref{eq:chiN_det}), we obtain the generating function in the
form of a ratio of two determinants,
\begin{align}\label{eq:chiN_det_non}
  \chi_N(\lambda) &= \frac{ \det \langle f_m | 1 - \mathcal{T} + \mathcal{T} e^{i \lambda} |
  f_n \rangle}{\det \langle f_m | f_n \rangle} \nonumber\\
  &=\frac{\det (\mathsf{S}^f - \mathsf{T}^f + \mathsf{T}^f e^{i \lambda})}
  {\det\mathsf{S}^f}
\end{align}
with the two $N\times N$ matrices
\begin{equation}\label{eq:matrices}
  \mathsf{S}^f_{mn} = \langle f_m | f_n \rangle \quad \text{and} \quad
  \mathsf{T}^f_{mn} = \langle f_m | \mathcal{T} | f_n \rangle.
\end{equation}

\subsection{Invariance of Slater Determinants under Linear
Transformations}\label{sec:invariance}

It turns out that the expression (\ref{eq:chiN_det_non}) for the generating
function can be drastically simplified and rewritten in a generalized
binomial form. As a first step towards this goal, one has to realize that
an $N$-dimensional Hilbert space $H_N$, spanned by the single-particle
wave-functions $f_n (k)$, defines exactly one properly antisymmetrized
wave function, or, equivalently, there exists (up to a phase factor) only
one associated $N$-particle Slater determinant state. The antisymmetrized
$N$-particle state is thus a property of the Hilbert space $H_N$ and is
independent on the basis chosen.\cite{roothaan:51}

Consider, as a simple example, a two-particle Slater-determinant state
(in second-quantized notation) $|\Psi\rangle = a_2^\dagger a_1^\dagger
|0\rangle$, with the vacuum-state $|0\rangle$ and Fermionic operators
$a_{1,2}$. Defining the new operators $a_\pm= (a_1 \pm a_2)/\sqrt{2}$,
we easily see that the two-particle state
\begin{equation}\label{eq:invariance2}
  a_+^\dagger a_-^\dagger |0\rangle = \frac{1}{2} (a_1^\dagger +
  a_2^\dagger)(a_1^\dagger - a_2^\dagger) |0\rangle
  = a_2^\dagger a_1^\dagger |0\rangle = |\Psi \rangle
\end{equation}
remains unchanged. Consider then a general $N$-particle Slater determinant
state of the form Eq.~(\ref{eq:slater_det_non}). Transforming the basis
states $f_m(k)$ to new states $g_m(k)$ via the complex linear transformation
\begin{equation}\label{eq:linear}
  g_{m}(k) = \sum_{n} \mathsf{A}_{nm} f_n (k), \qquad \det \mathsf{A} \neq 0,
\end{equation}
the antisymmetric combination
\begin{equation}\label{eq:f_trans_g}
  \det g_m(k_n) = (\det \mathsf{A}) \, \det f_m(k_n)
\end{equation}
remains invariant up to the factor $\det \mathsf{A}$; here, we have used
the fact that the determinant of the product of two matrices is the product
of the individual determinants. Furthermore, the normalized $N$-particle
Slater-determinant states $\Psi^f$ and $\Psi^g$ obey the relation
\begin{equation}\label{eq:psif_psig}
  \Psi^g (k_1, \dots, k_N) = \text{sgn}(\det \mathsf{A}) \Psi^f (k_1,\dots,k_N)
\end{equation}
with $\text{sgn}(x) = x/|x|$. The only effect of adopting a new basis is
the appearance of an overall phase factor $\text{sgn}(\det \mathsf{A})$
which drops out of the characteristic function (\ref{eq:chiN}). Therefore,
the full counting statistics calculated in the bases $f$ and $g$ give
identical results.

\subsection{Diagonalization}

The above invariance can be used to simplify the calculation of the
full counting statistics. Furthermore, even without specification of the
(time-independent) scatterer, one can obtain valuable insights about the
structure of possible outcomes in the counting statistics. In particular,
it turns out that the most general full counting statistics for a
Slater-determinant state is given by a generalized binomial expression of
the form (\ref{eq:fcs_1}).

Let us first investigate how the invariance under linear transformations,
Eq.~(\ref{eq:linear}), manifests itself in the determinant formula
Eq.~(\ref{eq:chiN_det_non}). To this end, we note that any single-particle
matrix $\mathsf{B}$ of the form (\ref{eq:matrices}) transforms under the
linear transformation $\mathsf{A}$ of the basis functions according to
\begin{equation}\label{eq:trans_M}
  \mathsf{B}^g = \mathsf{A}^\dagger \mathsf{B}^f \mathsf{A}, \qquad
  \mathsf{B} = \mathsf{S}, \mathsf{T}.
\end{equation}
Since $\det (\mathsf{A} \mathsf{B})= \det \mathsf{A} \det \mathsf{B}$, we
find that the characteristic function $\chi_N$ (we define $\mathsf{X}^f
= \mathsf{S}^f - \mathsf{T}^f + \mathsf{T}^f e^{i\lambda}$) is
invariant under the change of basis,
\begin{equation}\label{eq:trans_chi}
  \chi_N = \frac{\det \mathsf{X}^f}{\det \mathsf{S}^f} 
  = \frac{|\det \mathsf{A}\,|^2 \det \mathsf{X}^f}
  {|\det \mathsf{A}\,|^2 \det \mathsf{S}^f}
  = \frac{\det \mathsf{X}^g}{\det \mathsf{S}^g}.
\end{equation}
This invariance can be exploited by going over to new orthogonal basis
functions $g_m(k)$ with an overlap matrix $\mathsf{S}^g_{mn} =\delta_{mn}$
and a transmission matrix assuming a diagonal form $\mathsf{T}^g_{mn} =
\tau_m \delta_{mn}$.  The possibility of simultaneous diagonalization of
the matrices $\mathsf{T}^g_{mn}$ and $\mathsf{S}^g_{mn}$ is a consequence
of the transformation law (\ref{eq:trans_M}) characteristic of bilinear
forms (as opposed to linear transformations $\mathsf{L}$ which transform
according to $\mathsf{L}^g = \mathsf{A}^{-1} \mathsf{L}^f \mathsf{A}$)
combined with the positivity of $\mathsf{S}^f$. The corresponding
eigenbasis $g_m$ and eigenvalues $\tau_m$ of $\mathsf{T}^g_{mn}$ can be
found by solving the generalized eigenvalue problem
\begin{equation}\label{eq:gen_diag}
  (\mathsf{T}^f - \tau_m \mathsf{S}^f) a_m =0
\end{equation}
with the normalization $a_m^\dagger \mathsf{S}^f a_m^{\vphantom\dagger}=1$.
\cite{familiar} The eigenvectors $a_m$ constitute the column vectors
of the transformation matrix $\mathsf{A} = (a_1, \dots, a_N)$. The
eigenvalues are given by the roots of the characteristic polynomial $\det
( \mathsf{T}^f - \tau\, \mathsf{S}^f) =0$. The full counting statistics,
Eq.~(\ref{eq:chiN_det_non}), written in the new basis $g_m(k)$ assumes
the generalized binomial form
\begin{equation}\label{eq:binomial}
  \chi_N(\lambda) = \prod_{m=1}^N (1- \tau_m + \tau_m e^{i\lambda}),
\end{equation}
where the determinant has been evaluated explicitly and the result depends
only on the eigenvalues $\tau_m$. The generalized eigenvalue problem can
be reduced to a normal one by rewriting the problem in a orthogonalized
basis $\phi_m(k)$, with $\mathsf{S}^\phi = \openone_N$, which can be
obtained by the Gram-Schmidt procedure or by setting $\phi_m(k) = \sum_{n}
[(\mathsf{S}^f)^{-1/2}]_{nm} f_n(k)$.

From the above, we see that the concrete form of the eigenvalue problem
(\ref{eq:gen_diag}) is basis dependent, whereas the eigenvalues and vectors
are simply a property of the transmission operator $\mathcal{T}$ operating
in the Hilbert space $H_N$ with the scalar product $\langle f | g \rangle$.
Indeed, it is possible to find the eigenvalues and eigenvectors in a basis
independent way using the positive definite quadratic forms $T(g)=\langle g
| \mathcal{T} | g \rangle$ and $S(g)= \langle g | g \rangle$, $g \in H_N$.
Representing the bilinear form $T(g)$ with fixed $S(g)=1$ as a polar plot
with $T(g)$ the radius and $g$ defining the direction in $H_N$, we obtain
an ellipsoid in $N$-dimensional space. The lengths of the main axes of this
ellipsoid then constitute the eigenvalues and the associated directions
the eigenvectors of the problem (\ref{eq:gen_diag}). \cite{courant} The
eigenvalues $\tau_m$ are constrained to the interval $[0,1]$ as $T(g)\geq 0$
and $T(g) \leq S(g)$ due to unitarity.

\subsection{Full Counting Statistics for Entangled States}\label{sec:entangled}

The above discussion has concentrated on incoming states described by
a single Slater determinant, i.e., nonentangled states with Slater rank
1. It is instructive to generalize this discussion to entangled states
involving a coherent superposition of Slater determinants. We start from
an incoming state of $N$ particles with Slater rank 2,
\begin{equation}\label{eq:}
  \Psi(\mathbf{k}) = \alpha \Psi^\text{I}(\mathbf{k})
  + \beta\Psi^\text{II}(\mathbf{k}) ,
\end{equation}
where $\Psi^\text{I} (\mathbf{k})$ and $\Psi^\text{II} (\mathbf{k})$ are
normalized $N$-particle Slater determinants describing particles incoming
from the left and made from single particle states $f^\text{I}_m(k)$
and $f^\text{II}_m(k)$, $m=1,\dots, N$; the complex numbers $\alpha$ and
$\beta$ have been chosen such as to make $\Psi (\mathbf{k})$ normalized. The
characteristic function for the full counting statistics (\ref{eq:chiN})
assumes the form
\begin{widetext}
  \begin{equation}\label{eq:chi_ent}
    \chi_N (\lambda) = \Bigl[ \prod_{m=1}^N \int \frac{dk_m}{2\pi} (1 -
    T_{k_m} + T_{k_m} e^{i\lambda}) \Bigr] \Bigl[ |\alpha|^2
    |\Psi^\text{I}(\mathbf{k})|^2 + |\beta|^2
    |\Psi^\text{II}(\mathbf{k})|^2 +
    2 \text{Re}\{\alpha \beta^* \Psi^\text{I}(\mathbf{k})
    \Psi^\text{II}(\mathbf{k})^*\} \Bigr],
  \end{equation}
\end{widetext}
where $\text{Re}$ denotes the real part. The first two terms reduce to
generating functions for simple Slater determinant states and we can write
\begin{align}\label{eq:chi_ent2}
  \chi_N(\lambda) &= |\alpha|^2
  \chi^\text{I}_N(\lambda) + |\beta|^2
  \chi^\text{II}_N(\lambda) \nonumber\\
  &\qquad+ \alpha \beta^*\,
  \chi^\text{mix}_N(\lambda) + \alpha^* \beta \, \chi^\text{mix}_N(-\lambda)^*
\end{align}
with
\begin{align}\label{eq:chi_ent_det}
  \chi_N^\text{I}(\lambda) &= \frac{\det (\mathsf{S}^{f^\text{I}}
  - \mathsf{T}^{f^\text{I}} + \mathsf{T}^{f^\text{I}} e^{i \lambda})}
  {\det \mathsf{S}^{f^\text{I}}}, \nonumber \\
  \chi_N^\text{II}(\lambda) &= \frac{\det (\mathsf{S}^{f^\text{II}} -
  \mathsf{T}^{f^\text{II}} + \mathsf{T}^{f^\text{II}} e^{i\lambda})}
  {\det \mathsf{S}^{f^\text{II}}}, \nonumber\\
  \chi_N^\text{mix}(\lambda) &= \frac{ \det (\mathsf{S}^\text{mix} -
  \mathsf{T}^\text{mix} + \mathsf{T}^\text{mix} e^{i\lambda})}
  { \sqrt{\det \mathsf{S}^{f^\text{I}} \mathsf{S}^{f^\text{II}}} }.
\end{align}
The matrices with superscripts $f^\text{I}$ and $f^\text{II}$ have been
defined in Eq.~(\ref{eq:matrices}), while the new Hermitian matrices with
a superscript `mix' are given by the mixed matrix elements
\begin{equation}\label{eq:matrices_ent}
  \mathsf{S}^\text{mix}_{mn} =
  \langle f^\text{II}_m | f^\text{I}_n \rangle, \qquad
  \mathsf{T}^\text{mix}_{mn} =
  \langle f^\text{II}_m | \mathcal{T} | f^\text{I}_n \rangle.
\end{equation}
The first two terms in (\ref{eq:chi_ent2}) can be diagonalized as before,
cf.\ (\ref{eq:gen_diag}),
\begin{align}\label{eq:chi_ent_diag}
  \chi_N^\text{I}(\lambda) &=
  \prod_{m=1}^N (1- \tau^\text{I}_m+ \tau^\text{I}_m
  e^{i\lambda}), \\
  \chi_N^\text{II}(\lambda)& =
  \prod_{m=1}^N (1- \tau^\text{II}_m+ \tau^\text{II}_m
  e^{i\lambda}),
\end{align}
with the eigenvalues $\tau^\text{I}_m$ and $\tau^\text{II}_m$ given by the
roots of $\det (\mathsf{T}^\text{I}- \tau^\text{I} \mathsf{S}^\text{I})=0$
and $\det (\mathsf{T}^\text{II}- \tau^\text{II} \mathsf{S}^\text{II})=0$.

Let us then concentrate on the characteristic function $\chi^\text{mix}
(\lambda)$.  Unfortunately, there is no generic procedure to follow in this
case, as the matrices $\mathsf{S}^\text{mix}$ and $\mathsf{T}^\text{mix}$
are not Hermitian any more and hence the expression (\ref{eq:chi_ent})
cannot be further simplified in general. In particular, the characteristic
function $\chi^\text{mix}(\lambda)$ is not invariant under individual
transformations of the bases $f^\text{I}_m$ and $f^\text{II}_m$ (such
basis transformations leave the Slater-determinants invariant only up to
a phase factor, which dropped out in the calculation of the characteristic
function of a single Slater determinant state but does not when two Slater
determinants are superimposed coherently). In order to proceed further,
we restrict ourselves to specific situations where $\mathsf{S}^\text{mix}
= 0$ or $\det\mathsf{S}^\text{mix} \neq 0$.  The most trivial case is
realized for mutually orthogonal sets of basis functions $f^\text{I}_m$
and $f^\text{II}_m$ where $\mathsf{S}^\text{mix} = 0$; if, in addition,
$\det\mathsf{T}^\text{mix} = 0$, we have $\chi^\text{mix}(\lambda)=0$
(see also Sec.~\ref{sec:ent} below), else $\chi^\text{mix}(\lambda)
= \tau^\text{mix} (e^{i\lambda} -1)^N$ with $\tau^\text{mix} =
\det\mathsf{T}^\text{mix}/ \sqrt{\det \mathsf{S}^{f^\text{I}}
\mathsf{S}^{f^\text{II}}}$.

Second, let us assume that $\mathsf{S}^\text{mix}$ is invertible, $\det
\mathsf{S}^\text{mix} \neq 0$.  Let $\tau^\text{mix}_m$ be the roots of
the polynomial
\begin{equation}\label{eq:poly}
  \det[\mathsf{T}^\text{mix}-\tau^\text{mix} \mathsf{S}^\text{mix}]=0.
\end{equation}
The matrix $\mathsf{T}^\text{mix}(\mathsf{S}^\text{mix})^{-1}$ then can
be brought into a Jordan canonical form with $\tau^\text{mix}_m$ on the
diagonal and the characteristic function assumes the simple form
\begin{equation}\label{eq:chi_ex}
  \chi^\text{mix}_N(\lambda) = \frac{\det \mathsf{S}^\text{mix}}{ \sqrt{\det
  \mathsf{S}^{f^\text{I}} \mathsf{S}^{f^\text{II}}}}
  \prod_{m=1}^N(1-\tau^\text{mix}_m+\tau^\text{mix}_m
  e^{i\lambda}).
\end{equation}
The procedure outlined above is straightforwardly generalized to states
with higher Slater rank.

\section{Two particles}\label{sec:two_particles}

\subsection{Full counting statistics}

The above findings have interesting generic consequences for the charge
transport of fermionic particles; in the following, we discuss the simplest
case of two particles, see Fig.~\ref{fig:setup}, where nontrivial exchange
properties manifest themselves.  For $N=2$ particles the diagonalization
(\ref{eq:gen_diag}) can be carried out explicitly for arbitrary matrices
$\mathsf{T}^f$ and $\mathsf{S}^f$. The two eigenvalues $\tau_{1,2}$ are
given by
\begin{equation}\label{eq:eig_2}
  \tau_{1,2} = \frac{\alpha \mp \sqrt{\alpha^2 
  - \det \mathsf{T}^f \det \mathsf{S}^f}}{\det \mathsf{S}^f},
\end{equation}
where the parameter $2\alpha\!=\!\mathsf{S}_{22}^f\!\mathsf{T}_{11}^f\!+
\!\mathsf{S}_{11}^f\!\mathsf{T}_{22}^f\!-\!2 \text{Re}(\mathsf{S}_{12}^f
\!\mathsf{T}_{21}^f)$. Alternatively, the eigenvalues $0\leq \tau_m \leq 1$
are given by a minimum/maximum property \cite{courant}
\begin{equation}\label{eq:minmax}
  \tau_1 = \min_{\makebox[12mm]{$\scriptstyle g \in H_2|
  S(g)=1$}} T(g), \qquad \tau_2 = \max_{\makebox[12mm]{$\scriptstyle g
  \in H_2|S(g)=1$}} T(g),
\end{equation}
with the eigenvectors $g_{1,2}(k)$ given by those functions where the
minimum/maximum values are attained, i.e, $T(g_{1,2}) = \tau_{1,2}$. Once
the eigenvalues $\tau_m$ are known, the characteristic function $\chi_2$
assumes the simple generalized binomial form
\begin{equation}\label{eq:diag_chi2}
  \chi_2(\lambda) = (1 - \tau_1 + \tau_1 e^{i \lambda})
  (1-\tau_2 +\tau_2 e^{i\lambda}).
\end{equation}
As a result, we find that in the new basis $g_m$, the two particles
traverse the scatterer independent of one another, i.e., the characteristic
function is a simple product of independent one-particle characteristic
functions. Even more, the characteristic function is determined by the
Hilbert space spanned by the incoming states $f_{1,2}$ and is independent of
the choice of basis.  Exchange effects manifest themselves when comparing
the result (\ref{eq:diag_chi2}) for the Slater determinant $\Psi^f
\propto \det f_m(k_n)$ with the result $\chi_2^\text{dist}(\lambda) = (1 -
\mathsf{T}^f_{11} +\mathsf{T}^f_{11} e^{i \lambda})(1 - \mathsf{T}^f_{22}+
\mathsf{T}^f_{22} e^{i \lambda})$ for distinguishable particles,
$\Psi^\text{dist} \propto f_1(k_1) f_2(k_2)$: Exchange effects are absent
if both matrix elements $\mathsf{S}^f_{21}=\langle f_2|f_1\rangle=0$ and
$\mathsf{T}^f_{21} = 0$, i.e., for orthogonal initial and transmitted
states. On the other hand, a finite overlap of at least one pair of
these states generates finite exchange effects via the substitution of
$\mathsf{T}^f_{mm}$ in $\chi_2^\text{dist}$ by the eigenvalues $\tau_m$
in $\chi_2$.
\begin{figure*}[tb]
  \centering
  \includegraphics{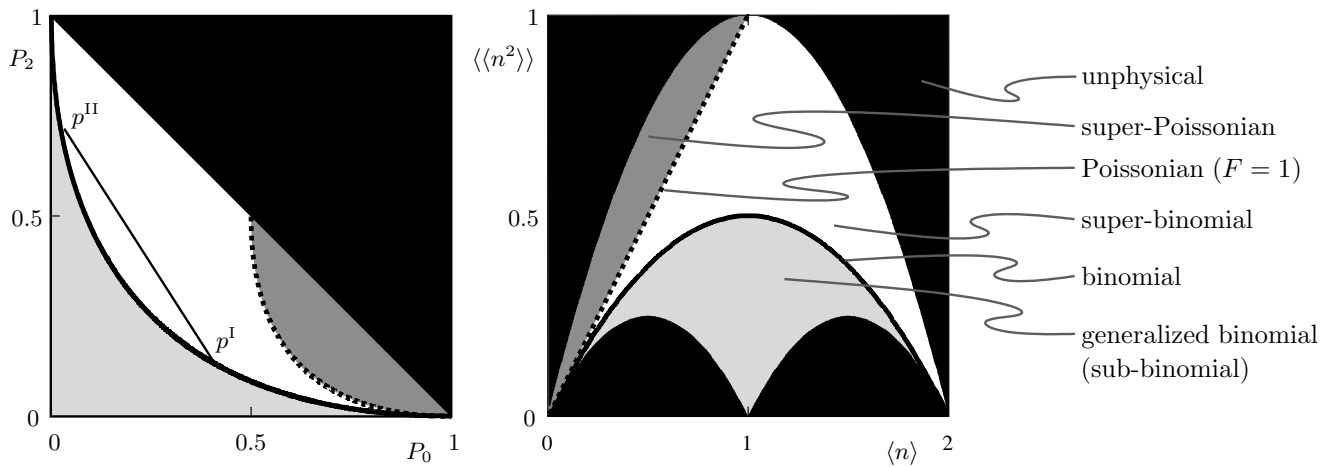}
  \caption{%
  Diagrams describing the generic statistical properties of two-particle
  transmission, on the left as a $P_2$-$P_0$ probability diagram, on the
  right as a noise-charge $\langle\langle n^2 \rangle\rangle$-$\langle n
  \rangle$ diagram.  The black regions are unphysical with probabilities
  $P_0,P_1,P_2$ residing outside $[0,1]$. The light gray regions describe
  generalized binomial (sub-binomial) processes, Eq.~(\ref{eq:diag_chi2}),
  bounded by the black line characterizing usual binomial processes. The
  dotted lines correspond to a Fano factor $F=\langle\langle n^2
  \rangle\rangle/\langle n \rangle$ equal to one. Within the dark gray
  regions noise is super-Poissonian with a Fano factor $F>1$.  Note that
  in order to observe super-Poissonian noise the reflection probability
  has to be large, such that $P_0 >1/2$ and $\langle n \rangle<1$.
  }
  \label{fig:cond}
\end{figure*}

The minimum/maximum property described above entails a set of \emph{a
priori} inequalities for the transmission probabilities $P_n$ involving
the transmission matrix elements $T_\text{min} = \min\{\mathsf{T}_{11}^f,
\mathsf{T}_{22}^f\}$ and $T_\text{max} = \max\{\mathsf{T}_{11}^f,
\mathsf{T}_{22}^f\}$; note that while the probabilities $P_n$ do account for
exchange effects, the single particle matrix elements $\mathsf{T}_{mm}^f$
obviously do not. With initial (nonorthogonal) wave packets $f_m$ normalized
to unity, $S(f_m)=1$, the search for the extrema in Eq.~(\ref{eq:minmax})
includes these states as well. We then obtain the set of inequalities $0 \leq
\tau_\text{1} \leq T_\text{min} \leq T_\text{max} \leq \tau_\text{2} \leq
1$.  Using them to estimate $P_0 = (1-\tau_1)(1-\tau_2) \leq (1-\tau_2)$,
$P_2=\tau_1 \tau_2 \leq \tau_1$, and $P_1 =1-P_0-P_2$, we can derive the
following bounds
\begin{equation}\label{eq:bound}
  P_0 \leq 1- T_\text{max}, \quad P_1 \geq T_\text{max}- T_\text{min},
  \quad P_2 \leq T_\text{min}.
\end{equation}
for the {\it transmission probabilities for two particles}. The above bounds
set an upper limit on bunching ($P_2$ and $P_0$) and a lower limit on
anti-bunching ($P_1$). Note though, that the bound on $P_2$ does not exclude
an increase (due to exchange) of the transmission probability beyond the
`classical' value $P_2^\text{dist} =\mathsf{T}^f_{11} \mathsf{T}^f_{22}$
for distinguishable particles, see $\chi_2^\text{dist}$ above. Indeed,
since $\mathsf{T}^f_{11} \mathsf{T}^f_{22} \leq T_\text{min}$, a value $P_2
\gg \mathsf{T}^f_{11} \mathsf{T}^f_{22}$ remains possible. Such a result
has been recently observed:\cite{lebedev:08} the probability of two-electron
events in the electron emission from a $\text{Cs}_3\text{Sb}$ photocathode
in a photomultiplier tube has been found to be much larger than the square
of the probability for single-electron emission.  This was observed both
in the case of thermal emission without photocathode illumination and
photoemission under weak photocathode illumination.  Furthermore, as
detailed calculation shows, a large $P_2$ can also be obtained for wave
packets with amplitudes $f_2(k) = f_1(k+\delta k)$ shifted in $k$-space and
a large overlap integral $\mathsf{S}^f_{21}$, combined with a transmission
amplitude suppressing $k$-values in the overlap region.

\subsection{Restrictions due to binomial statistics}

An arbitrary two-particle scattering process is fully characterized by the
three parameters $P_0, P_1, P_2$, from which only two are independent;
here, we assume that we can transmit only integer charges (no charge
fractionalization). In Figs.~\ref{fig:cond}(a) and (b), we find the
regions with different statistical properties that can be generated in a
two-fermion scattering process, both in $P_0$-$P_2$ parameter space as well
as in the noise $\langle\langle n^2\rangle\rangle$ versus average number
$\langle n\rangle$ diagram. We start with the definition of the physically
accessible regime in these diagrams: requiring that $P_1=1-P_0-P_2\geq 0$
(Fig.~\ref{fig:cond}(a)) and $P_0,~P_1,~P_2 \geq 0$ (Fig.~\ref{fig:cond}(b)),
we find that the black regions are forbidden.

Traditionally, starting from Poissonian statistics ($F=1$) relevant
for the coherent light emitted from a laser or for the transport of
a classical electron gas in a vacuum tube, much emphasis has been put
on the distinction between sub- and super-Poissonian statistics, with
reduced and enhanced noise intensity as quantified by Fano factors $F<1$
(sub-Poissonian noise) and $F>1$ (super-Poissonian processes). It appears to
us that in the context of degenerate fermions, the generic starting point is
the binomial statistics, instead, and more relevant qualifications are given
by the regimes of sub-binomial and super-binomial processes introduced below.

Nevertheless, let us start our analysis with the traditional classification
comparing a process with Poissonian statistics, which is realized on the
dotted line in Fig.~\ref{fig:cond}(a) defined through the relation
\begin{equation}\label{eq:fano_factor}
  F= \frac{P_0(1-P_0) + P_2 (1-P_2) + 2 P_0 P_2}{1-P_0+ P_2}=1,
\end{equation} 
i.e.,
\begin{equation}\label{eq:ineq_fano}
  P_2 = P_0 - \sqrt{2 P_0 -1 } \quad \text{with} \quad P_0 \geq 1/2.
\end{equation} 
Within the dark-gray region noise is super-Poissonian, which is usually
associated with the bunching of particles and therefore with bosonic
statistics.  Note that Fano-factors larger than the Poissonian value 1
require a large reflection probability $P_0>1/2$; only when most of the
particles are reflected one can observe the `bunching' of the remaining
transmitted objects.

A much more natural classification for our fermion system is in terms
of (deviations from) binomial statistics.  The characteristic function
$\chi_2$ for two fermions in a Slater determinant state can be cast into
the generalized binomial form Eq.~(\ref{eq:diag_chi2}), which depends
on two parameters $\tau_1,\tau_2$. As a consequence, the probabilities
satisfy the additional inequality
\begin{equation}\label{eq:ineq_binom}
  \sqrt{P_0} + \sqrt{P_2} \leq 1.
\end{equation} 
This condition follows from expressing the parameters $\tau_1,\tau_2$
through the probabilities $P_0, P_2$ using the relations $P_0 = (1-
\tau_1)(1-\tau_2)$ and $P_2 = \tau_1 \tau_2$; requiring a positive
discriminant of the resulting (quadratic) equation implies the
constraint (\ref{eq:ineq_binom}) which defines the light gray region in
Fig.~\ref{fig:cond}(a)), naturally termed the `sub-binomial' regime. The
(thick) black line bounding the general binomial (or sub-binomial) region
is the line of usual binomial statistics, which is realized for the case
of degenerate transmission coefficients $\tau_1=\tau_2$ as they appear if
the scattering does not depend on energy.

The region with super-Poissonian noise (dark gray) and the sub-binomial
region (light gray) are distinct, with the statistics of fermions incoming
in a Slater-determinant state always residing in the sub-binomial domain.
Note that the counting statistics of an arbitrary two-particle process
(without specification of exchange properties) also depends on two out of the
three parameters $P_0,P_1,P_2$ (as the constraint $P_0+P_1+P_2=1$ needs to
be fulfilled) but cannot be cast into the form Eq.~(\ref{eq:diag_chi2}) in
general, hence these processes are devoid of such an additional restriction.

The $P_2$-$P_0$ diagram can be transcribed to the (experimentally more
relevant) $\langle\langle n^2 \rangle \rangle$-$\langle n \rangle$ diagram,
cf.\ Fig.~\ref{fig:cond}(b). The physical constraints $0\leq P_0,P_1,P_2$
lead to the set of inequalities,
\begin{align}\label{eq:ineq_nn2}
  \langle\langle n^2 \rangle \rangle 
       &\geq \langle n \rangle (1- \langle n \rangle), \nonumber \\ 
  \langle\langle n^2 \rangle \rangle 
       &\leq \langle n \rangle (2- \langle n \rangle), \nonumber \\ 
  \langle\langle n^2 \rangle \rangle 
       &\geq (\langle n \rangle-1) (2- \langle n \rangle),
\end{align} 
which can be cast into the more compact form $(m +1 -\langle n
\rangle)(\langle n \rangle - m) \leq \langle\langle n^2 \rangle\rangle \leq
\langle n \rangle (2 - \langle n \rangle)$, with $m=0,1$. The single large-
and two small parabolas bounding the unphysical (black) regions are given by
the second and the two (for $m=0,1$) first inequalities. For the generalized
(or sub-) binomial statistics, the additional constraint assumes the form
\begin{equation}\label{eq:fano2}
  F =\frac{\langle\langle n^2 \rangle\rangle}{\langle n\rangle} \leq
  1 -\langle n\rangle/2,
\end{equation} 
with the equality applying to the binomial case with $\tau_1=\tau_2$.
Within the gray region of the diagram the noise is sub-binomial $F \leq
1- \langle n\rangle/2$ and hence trivially sub-Poissonian, $F \leq 1$.
Note that noiseless transmission of charge requires that an integer average
charge is transmitted.

The generalization of the above analysis to $N$ incoming particles in
a Slater determinant state is straightforward.  The generalized binomial
characteristic function is given by Eq.~(\ref{eq:fcs_1}). The positivity of
the probabilities $P_m \geq 0$, $m=0,\dots,N$ imposes the $N+1$ restrictions
on the first two momenta $\langle n \rangle$ and $\langle\langle n^2
\rangle\rangle$, $(m +1 -\langle n \rangle)(\langle n \rangle - m) \leq
\langle\langle n^2 \rangle\rangle \leq \langle n \rangle (N - \langle
n \rangle)$, with $m=0,\dots,N-1$, defining a simple generalization of
Fig.~\ref{fig:cond}(b) with one large and $N$ small parabolas. In the
generalized binomial case, the additional constraint
\begin{equation}\label{eq:fano_n}
  F = \frac{\langle\langle n^2 \rangle\rangle}{ \langle n\rangle} \leq
  1 -\langle n\rangle/N\leq 1
\end{equation} 
tells that the incoming Slater determinant states produce a sub-binomial noise
statistics. A similar result was found recently\cite{abanov:08} in the context
of adiabatic pumping. The authors considered a time-dependent scattering
matrix in the instant scattering approximation (i.e., an energy independent
scatterer) and obtained a generating function in a product form describing a
generalized binomial statistics with parameters $u_m \leq 0$; the $u_m$ relate
to our $\tau_m$ via $\tau_m = (1-u_m)^{-1}$.

\subsection{Entangled states}\label{sec:ent}

The above discussion for two particles lets us conclude that incoming Slater
determinant states generate Fano factors $F\leq 1-\langle n \rangle/2
\leq 1$; such states are nonentangled. On the other hand, an entangled
two-particle state can be generated with a sum of two Slater determinants;
such an entangled state (with $0<\alpha<1$, i.e., a state with Slater rank
2, see Sec.~\ref{sec:entangled})
\begin{equation}\label{eq:wf_ent}
  \Psi(k_1,k_2) = \sqrt{\alpha}\Psi^{I}(k_1,k_2) +
  \sqrt{1-\alpha} \Psi^\text{II}(k_1,k_2),
\end{equation}
is sufficient to generate all possible types of two-particle statistics:
we choose the Slater-determinant wave functions $\Psi^\text{I}$ and
$\Psi^\text{II}$ (incoming from the left) such that they occupy different
parts of momentum space, e.g., $\Psi^\text{I}$ has only components below
$k_\text{c}$ and $\Psi^\text{II}$ above. Furthermore, let the transmission
be $T_1= T_{k< k_\text{c}}$ below $k_\text{c}$ and $T_2= T_{k>k_\text{c}}$
above.  For such a setup, all the overlap integrals vanish, e.g., $\int
(dk_1 dk_2/4\pi^2) \Psi^{\text{II}*} (k_1,k_2) \Psi^\text{I}(k_1,k_2) =
0$, and we obtain (cf.\ Eq.~(\ref{eq:chi_ent2}))
\begin{equation}\label{eq:chi2_ent}
  \chi_2(\lambda)\!=\!\alpha (1 - T_1 + T_1 e^{i \lambda})^2 + (1\!-\!\alpha)
  (1 - T_2 + T_2 e^{i \lambda})^2,
\end{equation}
that is, the generating function is simply the weighted sum of the two
individual generating functions for the Slater-determinant states. The
statistics of such entangled wave functions is described by points in the
$P_2$-$P_0$ diagram of Fig.~\ref{fig:cond}(a) which lie on a straight
line between the point $p^\text{I}$ for $\Psi^\text{I}$ and the point
$p^\text{II}$ for $\Psi^\text{II}$ with $\alpha$ parameterizing the
line.  Both $p^\text{I}$ and $p^\text{II}$ are situated on the binomial
line, while the line connecting them may enter the super-binomial or
even the super-Poissonian region: for example setting $T_1 = 0$ and
$T_2=1$, the characteristic function is given by $\chi_2 = \alpha + (1-
\alpha) e^{2i\lambda}$ and $F=2\alpha$, which assumes values between
zero and two (note that in the limit $\alpha\to 1$, the wave function
(\ref{eq:wf_ent}) is of Slater rank 1, but nevertheless, the Fano factor
approaches $F=2$. As the Fano factor for $P_0=1$ assumes the form 0/0
its value depends on the direction from which $P_0=1$ is approached). As
simple Slater determinants produce only Fano factors up to $1-\langle
n \rangle/2$, a larger value serves as a test for the entanglement
of the two particles.\cite{dejong:94,bodoky:08}.  For $N$ incoming
particles in an entangled state of rank 2, the analogous construction
(cf.~\ref{sec:entangled}) produces a Fano factor $F= N \alpha$ with
$0< \alpha <1$, i.e., super-Poissonian statistics can be admitted for
sufficiently large $\alpha$.

\subsection{Two spin $1/2$ particles}\label{sec:spin}

Next, we consider the situation in the setup of Fig.~\ref{fig:setup} with
incoming particles in normalized states $f_1(k)$ and $f_2(k)$ with overlap
$S=\mathsf{S}^f_{21} =\langle f_2 | f_1 \rangle$ and carrying a spin $1/2$
degree of freedom. We consider the case of spin-independent scattering,
hence the coefficients in $\mathsf{T}^f$ depend exclusively on $f_1(k)$
and $f_2(k)$. The four properly symmetrized states available to the two
incoming particles are denoted by $\Psi_{s,m_s} (\mathsf{k})$, with $s=0$
the singlet ($m_s=0$) state and $s=1$ the three ($m_s=-1,0,+1$) triplet
states. The degrees of freedom $\mathbf{k}$ involve the momenta $k_m$
and spins $s_m$ of the particles, $\mathbf{k}=(k_1,s_1;k_2,s_2)$. The
triplet states with $m_s=\pm1$ are simple Slater determinant states
\begin{align}\label{eq:triplet}
  \Psi_{1,\pm 1}(\mathbf{k}) &= \frac{1}{\sqrt{2(1-|S|^2)}} \bigl[f_1(k_1)
  \chi_{\uparrow/\downarrow}(s_1) f_2(k_2)  \chi_{\uparrow/\downarrow}(s_2)
  \nonumber\\
  & \qquad\quad - [(k_1,s_1) \leftrightarrow (k_2,s_2)]  \bigr].
\end{align}
The characteristic function $\chi_2$ for the full counting statistics
then is of the generalized binomial form with $\tau_{1/2}$ given by
Eq.~(\ref{eq:eig_2}),
\begin{align}\label{eq:fcs_triplet}
  \chi_{1,\pm 1}(\lambda)&= (1-\tau_1 + \tau_1 e^{i \lambda}) (1-\tau_2 +
  \tau_2 e^{i \lambda})\\
  &=
  \frac{ (1-\mathsf{T}^f_{11} + \mathsf{T}^f_{11} e^{i\lambda})
    (1-\mathsf{T}^f_{22} + \mathsf{T}^f_{22} e^{i\lambda})}{1- |S|^2}
  \nonumber\\
   &\quad - \frac{(S - \mathsf{T}^f_{21}+\mathsf{T}^f_{21} e^{i\lambda})
   (S^* - \mathsf{T}^f_{12}+\mathsf{T}^f_{12} e^{i\lambda})}{1-|S|^2}. \nonumber
\end{align}

The states with $m_s=0$ are more interesting as they are of Slater rank 2.
Defining
\begin{align}\label{eq:fi/ii}
  f^\text{I}_1(k,s)
  &= f_1(k) \chi_\uparrow (s),
  &f^\text{I}_2(k,s)
  &= f_2(k) \chi_\downarrow (s), \nonumber\\
  f^\text{II}_1(k,s)
  &= f_1(k) \chi_\downarrow (s),
  &f^\text{II}_2(k,s) &= f_2(k)
  \chi_\uparrow(s),
\end{align}
we have
\begin{equation}\label{eq:singlet}
  \Psi_{0/1,0}(\mathbf{k}) = \frac{1}{\sqrt {2 (1 \pm |S|^2)} }
  [\Psi^\text{I} (\mathbf{k})
  \mp \Psi^\text{II} (\mathbf{k}) ]
\end{equation}
with $\Psi^\text{I/II}(\mathbf{k})$ the normalized two-particle Slater
determinants made from the states $f^\text{I/II}_m$. The calculation of
the characteristic function follows the procedure outlined above: As the
matrices $\mathsf{T}^\text{I/II}_{mn}=\mathsf{T}^f_{mm} \delta_{mn}$ and
$\mathsf{S}^\text{I/II}_{mn}=\delta_{mn}$ are diagonal (the particles 1
and 2 are distinguishable), we immediately have
\begin{equation}\label{eq:fcs_i/ii}
  \chi^\text{I/II}(\lambda) =
  (1-\mathsf{T}^f_{11} + \mathsf{T}^f_{11}
  e^{i\lambda}) (1-\mathsf{T}^f_{22} + \mathsf{T}^f_{22}
    e^{i\lambda}).
\end{equation}
For the calculation of $\chi^\text{mix}(\lambda)$, the matrices
$\mathsf{T}^\text{mix}$ and $\mathsf{S}^\text{mix}$ need to be evaluated. In
the present case, they are purely off-diagonal with the off-diagonal
matrix element given by $\mathsf{T}^\text{mix}_{21}=\mathsf{T}^f_{21}$
and $\mathsf{S}^\text{mix}_{21} = S$. Calculating the determinants in
Eq.~(\ref{eq:chi_ent_det}), we obtain the mixed component in the form
\begin{equation}\label{eq:fcs_mix}
  \chi^\text{mix}(\lambda) = -(S - \mathsf{T}^f_{21}
  + \mathsf{T}^f_{21} e^{i\lambda})
  (S^*
  - \mathsf{T}^f_{12} + \mathsf{T}^f_{12} e^{i\lambda})
\end{equation}
and the characteristic function is given by
\begin{multline}\label{eq:fcs_singlet}
  \chi_{0/1,0}(\lambda) =
  \frac{ (1-\mathsf{T}^f_{11} + \mathsf{T}^f_{11} e^{i\lambda})
    (1-\mathsf{T}^f_{22} + \mathsf{T}^f_{22} e^{i\lambda})}{1\pm |S|^2}
  \\
   \pm \frac{(S - \mathsf{T}^f_{21}+\mathsf{T}^f_{21} e^{i\lambda})
   (S^* - \mathsf{T}^f_{12}+\mathsf{T}^f_{12} e^{i\lambda})}{1\pm|S|^2}.
\end{multline}
The result (\ref{eq:fcs_singlet}) agrees with the results in
Ref.~\onlinecite{hassler:07}. The characteristic functions for the
two spin triplet states $s=1$ with maximal magnetization $m_s=\pm1$
and the characteristic function for the triplet $s=1,m_s=0$ with zero
magnetization coincide with the one for a Slater determinant of spinless
fermions, Eq.~(\ref{eq:diag_chi2}). This is because all three states involve
identical orbital wave functions and the scattering process does not depend
on the spin part of the wave function.  The corresponding average number
of particles $\langle n \rangle_{1,m_s}$ and noise $\langle\langle n^2
\rangle\rangle_{1,m_s}$ reside within the region of generalized binomial
statistics, cf.\ Fig.~\ref{fig:cond},
\begin{equation}\label{eq:ineq_singlet}
  F_{1,m_s} \leq 1-\langle n\rangle_{1,m_s}/2.
\end{equation}
The entangled singlet state (with $s=0$) does not necessarily fulfill this
condition. Rather opposite, for the case where the individual transmission
probabilities of the two particles are equal, $\mathsf{T}^f_{11} =
\mathsf{T}^f_{22}$, the moments and the Fano factor always reside outside
the region allowed by the generalized binomial statistics,
\begin{equation}\label{eq:ineq_triplet}
   F_{0,0} \geq 1-\langle n\rangle_{0,0}/2,
\end{equation}
as a lengthy but straightforward calculation shows. Hence, this rather
trivial setup can be used to discriminate singlet from triplet states and
also serves as an indicator of entanglement (as long as the inequalities
are strict which is the case as long as $S\neq 0$ and $\mathsf{T}^f_{21}
\neq S \mathsf{T}^f_{11}$).

A similar experiment was proposed by Burkard \emph{et al.}
Ref.~\onlinecite{burkard:00}, which had two particles with equal
energy come in from different arms in a symmetric beam splitter, see
Ref.~\onlinecite{taddei:02} for a calculation of the full counting statistics
for this setup. Our setup involves one single lead only, at the expense of
requiring an energy dependent transmission probability (otherwise we end up
on the binomial line which is devoid of any separation power). Furthermore,
the discrimination between singlet and triplet states is determined by the
presence or absence of generalized binomial statistics and hence involves
the binomial bound $1-\langle n\rangle_{0,0}/2$ on $F$.

\section{$N$-particle trains} \label{sec:train}

We consider the case of $N$ incoming particles, all with the same shape of
the wave function $f(k)$ aligned regularly in real space with separation $a$;
the wave function of the $m$-th particle then is given by $f_m(k) = f(k)
e^{-i m a k}$.  The overlap and transmission matrices (\ref{eq:matrices})
are given by the Fourier transforms
\begin{align}\label{eq:overlap}
  \mathsf{S}^f_{mn} &= \int \frac{dk}{2\pi}\, |f(k)|^2 e^{i (m-n) k a},
  \nonumber\\ \mathsf{T}^f_{mn} &= \int \frac{dk}{2\pi}\, |f(k)|^2 T_k 
  e^{i (m-n) k a};
\end{align}
these are Toeplitz matrices as their elements depend only on the difference
$m-n$ between indices. In the limit $N\to \infty$, the determinants of the
Toeplitz matrices $\mathsf{S}^f$ and $\mathsf{S}^f-\mathsf{T}^f+\mathsf{T}^f
e^{i\lambda}$ can be evaluated by reducing the integral over $k$-space to
an integral over the first Brillouin zone $[0,2\pi/a]$ and using Szeg\H{o}'s
theorem, see Ref.~\onlinecite{szego:15} and App.~\ref{sec:szego},
\begin{widetext}
  \begin{align}\label{eq:szego}
    \log \det \mathsf{S}^f &\sim N \int_0^{2\pi}\!\! \frac{d \theta}{2\pi}
    \log \Bigl\{ \frac{1}{a}\sum_{m \in \mathbb{Z}} | f[(\theta + 2\pi m) /a ]|^2
    \Bigr\} , \nonumber\\
    \log \det (\mathsf{S}^f - \mathsf{T}^f + \mathsf{T}^f e^{i\lambda}) &\sim N
    \int_0^{2\pi}\!\! \frac{d\theta}{2\pi} \log \Bigl\{ \frac{1}{a} 
    \sum_{m \in \mathbb{Z}} 
    | f[(\theta + 2\pi m) /a ]|^2 [1- T_{(\theta + 2\pi m) /a} +T_{(\theta +
    2\pi m) /a} e^{i \lambda}] \Bigr\}.
  \end{align}
\end{widetext}
The logarithm of these determinants scales linearly with $N$, a result
that has to be expected as correlations between particles vanish at
large separation.  Combining the results (\ref{eq:szego}) and replacing
the integration over the angle $\theta$ by an integration over the first
Brillouin zone $k \in [0,2\pi/a]$, we find the generating function in
the form
\begin{equation}\label{eq:char}
  \log \chi_N
  (\lambda) = N a \int_0^{2\pi/a} \! \frac{dk}{2\pi} \log(
  1 - \tau_k + \tau_k e^{i \lambda})
\end{equation}
with the effective scattering probabilities
\begin{equation}\label{eq:eff_scatter}
  \tau_k = \frac{ \sum_{m \in \mathbb{Z}} | f(k + 2 \pi m/a) |^2 T_{k+2
  \pi m/a} } { \sum_{m \in \mathbb{Z}} | f(k + 2\pi m/a)|^2 },
\end{equation}
which denote transmission probabilities (with $0\leq \tau_k \leq 1$)
averaged over higher harmonics $2\pi m/a$ with weight $|f(k+2\pi m/a)|^2$.

Let us apply this result to wave packets generated by Lorentzian voltage
pulses. As shown in Ref.~\onlinecite{keeling:06}, a unit-flux (i.e., $c\int
dt \, V(t) = hc/e = \Phi_0$) Lorentzian voltage-pulse $e V_{t_0} (t) = 2
\hbar \gamma / [(t-t_0)^2 + \gamma^2]$, parametrized by its width $\gamma$
and time of appearance $t_0$, excites a single particle with wave function
$f_{x_0}(k)= \sqrt{4\pi \xi} e^{- \xi k - i x_0 k} \Theta(k)$ moving through
the quantum wire ($\Theta(k)$ denotes the unit-step function; we remind
that $k$ is measured with respect to the Fermi momentum $k_{\F}$). Here
$e$ is the charge of the particle, $x_0 = v_{\F} t_0$ parametrizes the
position, and $\xi = v_{\F} \gamma$ the real-space width of the wave
packet. A periodic sequence of unit-flux voltage-pulses $V(t) = \sum_{m
\in \mathbb{Z}} V_{m a/v_{\F}}(t)$ applied to an interval to the left
of the scatterer and driving one particle per time interval $a/v_{\F}$
generates the transmission probabilities
\begin{equation}\label{eq:voltage_pulses}
  \tau_k = (1 - e^{- 4 \pi \xi/a}) \sum_{m \geq 0} e^{-4 \pi m \xi/a} T_{k+
  2\pi m/a}.
\end{equation}
\begin{figure}[t]
  \centering
  \includegraphics{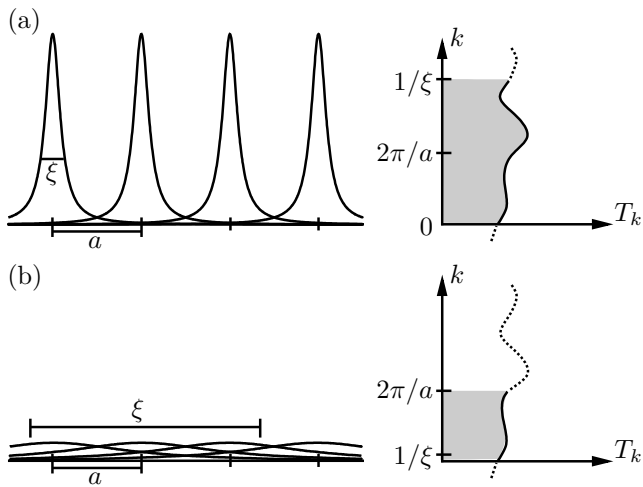}
  \caption{%
  (a) Train of nonoverlapping wave-packets, $\xi\ll a$. Each particle is
  transmitted independent of the others with a transmission probability $\int
  (dk/2\pi) |f(k)|^2 T_k$ depending only on its momentum distribution $f(k)$.
  For wave-packets with width $\xi$, these probe transmission probabilities
  for momenta up to $1/\xi$. (b) Train of strongly-overlapping wave-packets,
  $\xi\gg a$. If the particles were transmitted independent of each other
  (no exchange effects), they would probe transmission probabilities
  $T_k$ in the range up to $1/\xi \approx 0$. Due to exchange effects,
  the particles fill up a Fermi sea determined by the density $1/a$.
  Therefore, the particle train probes transmission probabilities for
  momenta in the interval $[0,2\pi/a]$. This (stationary) state can be
  seen as a wave-packet analogue of the constant-voltage setup.
  }\label{fig:train}
\end{figure}
For nonoverlapping wave packets $\xi \ll a$, cf.\ Fig.~\ref{fig:train}(a),
exchange effects are absent. The sum in Eq.~(\ref{eq:voltage_pulses})
becomes an integral and the transmission probabilities assume the simple form
$\tau_k = \int (dk'/2\pi) |f(k')|^2 T_{k'} = \langle \mathcal{T}\rangle$
(as easily obtained from (\ref{eq:eff_scatter}) by replacing the sums with
integrals), independent of $k$: Every particle probes the transmission
probabilities with its weight $|f(k)|^2$.  In the opposite limit, $\xi \gg
a$, i.e., for flat wave-packets which are strongly overlapping, the result
for distinguishable particles is $\tau^\text{dist}_k \to T_0$. However,
exchange effects force the system to fill the first Brillouin zone $k \in
[0,2\pi/a]$ and the particles probe the transmission within this energy
interval $\tau_k = T_k$ (as obtained from (\ref{eq:voltage_pulses}) retaining
only the $m=0$ term), cf.\ Fig.~\ref{fig:train}(b).  Taking the limit $\xi
\to \infty$ corresponds to the case of a constant applied voltage $V$ of
magnitude $eV= 2\pi\hbar v_{\F}/a$ (we remind that the time integral over
one voltage pulse generates one flux unit $\Phi_0 = hc/e$ and pulses are
separated in time by $a/v_{\F}$) and the generating function assumes the form
\begin{equation}\label{eq:const_voltage}
  \log \chi_{N} (\lambda) \! = \! N \frac{2\pi\hbar v_{\F}}{eV} \!\!
  \int_0^{e V/\hbar v_{\F}} \!\!  
  \frac{dk}{2\pi} \log(1- T_k + T_k e^{i \lambda}).
\end{equation}
This result then is the characteristic function for the full counting
statistics for a constant voltage $V$ applied to the left of the scatterer
including an energy dependent scatterer. The equivalent result has been
found in Ref.~\onlinecite{schonhammer:07} if we perform the {\it ad hoc}
replacement of the particle number $N$ by a `measuring time' $t$, $N \to t
v_{\F}/a = t eV/2\pi\hbar$ (note, although this replacement appears sensible,
it is non rigorous as we have assumed the limit $t \to \infty$ within the
present scattering formalism; we will further comment on this later).

The next order term in the asymptotic expansion for $N\to\infty$ can be
obtained using the generalization of Szeg\H{o}'s theorem (Fisher-Hartwig
conjecture, cf.\ Eq.~(\ref{eq:log})) and is given by
\begin{equation}\label{eq:delta_chi_N}
  \Delta \log\chi_N(\lambda) = \frac{\log N}{4 \pi^2} \log^2 \Biggl[ \frac{ 1-
  T_{2\pi/a} + T_{2\pi/a} e^{i\lambda}} {1- T_0 + T_0 e^{i\lambda} } \Biggr];
\end{equation}
the logarithmic nature of this correction is due to the energy dependence of
the transmission coefficient $T_k$, in particular, its jump $T_0 \neq
T_{2\pi/a}$ across the first Brillouin zone $k \in [0,2\pi/a]$  (for $T_0 =
T_{2\pi/a}$ the correction term is of order unity, see (\ref{eq:szego_det})).
The correction for the noise term is given by
\begin{equation}\label{eq:delta_n2_N}
  \Delta \langle\langle n^2 \rangle \rangle = 
  \frac{ (T_{2\pi/a} - T_0)^2} {2\pi^2} \log N
\end{equation}
and similar corrections are obtained for the third- and higher-order
cumulants.

\section{Generalizations}\label{sec:generalizations}

\subsection{Unitary evolution and time-dependent counting}

We want to generalize the generating function $\chi_N$ as given by
Eq.~(\ref{eq:chiN_det}) to account both for the specific time-evolution of
the scattering state and for different counting procedures. Throughout this
discussion, it is convenient to apply the Dirac notation and we rewrite
the Slater determinant (\ref{eq:slater_det}) in the form
\begin{equation}\label{eq:init}
  |\Psi \rangle = \frac{1}{\sqrt{N!}} \sum_{\pi\in S_N} \text{sgn}(\pi)
  |\phi_{\pi(1)} \rangle \otimes \dots \otimes | \phi_{\pi(N)} \rangle;
\end{equation}
Eq.~(\ref{eq:init}) describes the initial $N$-particle wave function at time
$t=0$ composed of orthonormalized one-particle states $|\phi_m\rangle$
(here, $\pi$ denotes an element of the permutation group $S_N$). The
choice of orthonormalized wave packets is only for convenience: as seen
in Sec.~\ref{sec:invariance}, a Slater determinant is invariant under
general linear combination of states it is composed of; in particular,
an orthonormalized basis can be chosen.

Let
\begin{equation}\label{eq:evolution}
  \mathcal{U} = \exp \biggl[ -\frac{i}{\hbar}\int_{0}^t \!\! dt'\, 
  \mathcal{H}(t') \biggr]
\end{equation}
be the unitary evolution operator generated by the single particle
Hamiltonian $\mathcal{H}(t)$.\cite{timeorder} In the absence of interaction,
the evolution of the total system is governed by the product operator
$\Gamma_N(\mathcal{U})$, where, given a one-particle operator $\mathcal{O}$,
we define the $N$-particle operator
\begin{equation}\label{eq:tensor}
  \Gamma_N(\mathcal{O}) = \underbrace{\mathcal{O} \otimes \dots \otimes
  \mathcal{O}}_\text{$N$-times}
\end{equation}
acting simultaneously on all $N$-particles. While we restrict ourselves
to noninteracting systems, we still allow for a time-dependent scattering
potential which can generate inelastic processes. The final state at
time $t$ is given by $|\Psi_\text{out}\rangle = \Gamma_N(\mathcal{U})
|\Psi \rangle$. Including the counting field $e^{\pm i \lambda/2}$, the
wave-function assumes the form
\begin{equation}\label{eq:wave_pm}
  |\Psi_\text{out}^\pm \rangle = \Gamma_N(e^{\pm i \lambda \mathcal{Q}/2})
  |\Psi_\text{out}\rangle = \Gamma_N(e^{\pm i \lambda \mathcal{Q}/2} 
  \mathcal{U}) |\Psi \rangle,
\end{equation}
where $\mathcal{Q}$ is a projector ($\mathcal{Q}^2=\mathcal{Q}$ and
$\mathcal{Q}^\dagger = \mathcal{Q}$) on that part of the wave-function that
has been counted. E.g., in the original setup of Ref.~\onlinecite{lesovik:06}
with a spin at position $x_0$ and particles incoming from the left, the
operator $\mathcal{Q}_t = \int_I dx\, |x \rangle \langle x|$ projects onto
the causal interval $I=[x_0,x_0 +v_{\F} t]$ (no such operator $\mathcal{Q}$
mimicking a spin-counter can be defined for particles incident from both
sides); hence that part of the wave function which passed the counter during
the time $t$ picks up an additional phase $e^{\pm i \lambda/2}$.
Note that it is always the full phase $\lambda$ which is picked up, as
the particle is either measured (eigenvalue 1 of $\mathcal{Q}$) or not
(eigenvalue 0 of $\mathcal{Q}$).  The characteristic function of the full
counting statistics is given by the overlap (fidelity)
\begin{equation}\label{eq:chi}
  \chi_N(\lambda) = \langle \Psi_\text{out}^- | \Psi_\text{out}^+
  \rangle= \langle \Psi| \Gamma_N(\mathcal{U}^\dagger 
  e^{i\lambda \mathcal{Q} } \mathcal{U}) | \Psi \rangle
\end{equation}
of the forward- and back-propagating wave-functions measured with opposite
spin states.

Next, we exploit that the expectation value of a product operator
$\Gamma_N(\mathcal{O})$ in a Slater-determinant state can be written
as a determinant of one-particle matrix elements $\langle \phi_{m} |
\mathcal{O} | \phi_{n}\rangle$ in the Hilbert space $H_N$ spanned by the
states $|\phi_{m}\rangle$,
\begin{align}\label{eq:slater_to_det}
  \langle \Psi | \Gamma_N(\mathcal{O}) | \Psi \rangle &
  = \frac{1}{N!} \!\!\!\!\! \sum_{\pi,\pi' \in S_N} \!\!\!\!\!
  \text{sgn}(\pi \circ \pi')
  \prod_{m=1}^N \langle \phi_{\pi(m)} | \mathcal{O}| \phi_{\pi'(m)} \rangle
  \nonumber \\
  &=  \frac{1}{N!} \!\!\! \sum_{\pi,\pi'' \in S_N} \!\!\! 
  \text{sgn}(\pi'')
  \prod_{m=1}^N \langle \phi_{m} | \mathcal{O}| \phi_{\pi''(m)}
  \rangle
  \nonumber\\
  &= \det \langle \phi_{m} | \mathcal{O} | \phi_{n} \rangle;
\end{align}
this formula is at the origin of (most) results which cast the characteristic
function of the full counting statistics into a determinant form. Making
use of Eq.~(\ref{eq:slater_to_det}), we can rewrite the characteristic
function Eq.~(\ref{eq:chi}) as the determinant
\begin{align}\label{eq:determinant}
  \chi_N(\lambda) &= 
  \det \langle \phi_m | e^{i\lambda \mathcal{U}^\dagger \mathcal{Q}
  \mathcal{U}} | \phi_n
  \rangle \nonumber \\
  &=  \det \langle \phi_m | 1 -  
  \mathcal{T}_Q +  \mathcal{T}_Q e^{i\lambda}| \phi_n \rangle
\end{align}
with
\begin{equation}\label{eq:tq}
  \mathcal{T}_Q = \mathcal{U}^\dagger \mathcal{Q} \mathcal{U};
\end{equation}
in going from the first to the second line in Eq.~(\ref{eq:determinant}),
the exponential has been expanded and use has been made of the fact that
$\mathcal{T}_Q$ is a projector. With $\mathcal{T}_Q$ a projector in the
one particle Hilbert space $H$, its eigenvalues in the subspace $H_N$
lie between 0 and 1 and Eq.~(\ref{eq:determinant}) leads to a generalized
binomial statistics.

In order to familiarize us with this new formula, we reproduce the results
of the above section. We then are interested in the situation where the
initial state $|\Psi\rangle$ is localized to the left of the scattering
region and the final state describes the $t\to\infty$ asymptotic behavior
where all particles have completed the scattering process. Within the basis
of states left/right of the scatterer with momentum $k$, the asymptotic
form of the propagator is given by the unitary (scattering) matrix
\begin{equation}\label{eq:prop}
  U_k^\infty=
  \begin{pmatrix}
    r_k & t_k' \\
    t_k & r_k'
  \end{pmatrix},
\end{equation}
where the coefficients $r_k$ ($r_k'$) and $t_k$ ($t_k'$) are the reflection
and transmission amplitudes of a particle incoming from the left (right).
The total propagator assumes the form $\mathcal{U}^\infty= \int (dk/2\pi)
| k \rangle_\text{out} U_k^\infty \, {}_\text{in} \langle k|$ where
we have introduced the asymptotic states $|k \rangle_\text{in(out)}
= (|k \rangle_\text{L,in(out)}, |k \rangle_\text{R,in(out)})$
which are in-(out-)going plane waves in the left/right lead: in a
formal derivation, we have to consider the $t\to\infty$ limit of the
evolution in (\ref{eq:evolution}) within an interaction picture with
a trivial reference dynamics $\mathcal{U}_0 = \int (dk/2\pi) e^{-
i v_{\F} k t} (|k \rangle_\text{in} \, {}_\text{in}\langle k|+ | k
\rangle_\text{out} \, {}_\text{out}\langle k|)$. The counting operator
$\mathcal{Q}$ is given by the projection on the right outgoing lead,
$\mathcal{Q}_\text{R}=(0,1)^\dagger(0,1)$, and we obtain
\begin{equation}\label{eq:uqu}
  \mathcal{T}^\infty_{Q_\text{R}} = \int (dk/2\pi) | k \rangle_\text{in}
  (t_k , r'_k)^\dagger (t_k , r'_k)\, {}_\text{in} \langle k|.
\end{equation}
Since the initial single-particle wave functions ${}_\text{in}\langle
k|\phi_m\rangle = (\langle k | \phi_m \rangle,0)$ are located to the
left of the scatterer, the characteristic function assumes the form
\begin{equation}\label{eq:chit}
  \chi(\lambda) = \det \langle \phi_m 
  | 1 - \mathcal{T} + \mathcal{T} e^{i\lambda} | \phi_n \rangle,
\end{equation}
with $\mathcal{T} = \int (dk/2\pi)\,T_k |k\rangle\langle k|$, in agreement
with (\ref{eq:chiN_det}); here, we have shortened the notation $|k\rangle
= |k\rangle_\text{in,L}$ in agreement with previous sections. The
generalization of the result Eq.~(\ref{eq:chit}) to many channels is
straightforward: the propagator Eq.~(\ref{eq:prop}) exhibits a block
structure with matrices $t_k$ and $r_k$ describing the transmission
and reflection in the channel basis, the transmission probabilities
$T_k=t^\dagger_k t^{\vphantom\dagger}_k$ assume a matrix form and the state
vector $|\phi_m\rangle$ adopts an additional channel index. Assuming an
implicit summation over channel indices, the form of Eq.~(\ref{eq:chit})
remains unchanged. The same comment holds for the spin index.

\subsection{Density matrix -- finite temperatures}\label{sec:finite}

The determinant in Eq.~(\ref{eq:determinant}) is restricted to the subspace
spanned by the initial states $|\phi_m\rangle$. Introducing the projection
operator $\mathcal{P}=\sum_{m=1}^N |\phi_m \rangle \langle \phi_m |$ onto the
subspace spanned by the initial states $|\phi_m\rangle$, the determinant can
be elevated to cover the whole Hilbert space. We split the total Hilbert space
into the sector defined by the projector $\mathcal{P}$ and its complement projected
onto $\mathcal{P}_\perp = 1 - \mathcal{P}$. The operator $1-\mathcal{P}
\mathcal{T}_Q +  \mathcal{P} \mathcal{T}_Q e^{i\lambda}$ can be expressed
in block form
\begin{multline}\label{eq:block}
	[1+ \mathcal{P} \mathcal{T}_Q (e^{i\lambda}-1)] 
  \\
  =\begin{bmatrix}
    1+ \mathcal{T}_Q (e^{i\lambda}-1) &  \mathcal{T}_Q (e^{i\lambda}-1) \\
		0 &  1
  \end{bmatrix},
\end{multline}
with the blocks operating in the $\mathcal{P} H$ and $\mathcal{P}_\perp
H$ subspaces.  The determinant of the upper block-diagonal matrix
Eq.~(\ref{eq:block}) is given as the product of the determinant
(\ref{eq:determinant}) in the $P$-block and the determinant of $1$ in the
$P_\perp$-block and thus the generating function assumes the form
\begin{equation}\label{eq:determinant_general}
  \chi_N(\lambda) = \det(1- \mathcal{P} \mathcal{T}_Q 
	+ \mathcal{P} \mathcal{T}_Q e^{i\lambda}),
\end{equation}
where the determinant is taken over the entire one-particle Hilbert
space $H$.

Interestingly, this formula can be generalized to the case when the initial
state is not a single Slater determinant, but an incoherent superposition of
many Slater determinants with a density matrix of the form $\Gamma(\rho)/Z$
in Fock space $F = \bigoplus_N H_N$, $\Gamma(\mathcal{O}) = \bigoplus_N
\Gamma_N(\mathcal{O})$, $Z=\text{Tr}_{F_a} \Gamma(\rho)$, and $\rho$
is the one-particle density matrix, e.g., $\rho= e^{-\beta (\mathcal{H}
- \mu)}$ for a thermal ensemble with temperature $\beta^{-1}$, chemical
potential $\mu$ and time-independent {\it single}-particle Hamiltonian
$\mathcal{H}$; here $F_a$ denotes the antisymmetric sector of the
Fock space. Using the trace formula \cite{klich:03}, $\text{Tr}_{F_a}
[\Gamma(\mathcal{O})]= \det(1+\mathcal{O})$, where the determinant is over
the one-particle Hilbert space, the characteristic function $\chi(\lambda)
= \text{Tr}_{F_a} [\Gamma(\rho) \Gamma(\mathcal{U}^\dagger e^{i \lambda
\mathcal{Q}} \mathcal{U}) ]/Z$, cf.\ Eq.~(\ref{eq:chi}), assumes the form
\begin{align}\label{eq:finite_temp}
  \chi(\lambda) &= \det(1 + \rho e^{i \lambda \mathcal{U}^\dagger 
  \mathcal{Q} \mathcal{U}})/\det(1+\rho)
  \nonumber\\
  &= \det (1- \eta + \eta e^{i\lambda \mathcal{T}_Q}) \nonumber\\
	&= \det (1- \eta \mathcal{T}_Q + \eta \mathcal{T}_Q e^{i\lambda})
\end{align}
with the one-particle occupation-number operator $\eta=\rho/(1+\rho)$
(and arbitrary one-particle density matrix $\rho$); note again that
the spectrum of $\eta \mathcal{T}_Q$ resides between 0 and 1 so that
(\ref{eq:finite_temp}) denotes a generalized binomial statistics. 

As an example, consider the situation of two particles incident from the
left, with wave functions $\phi_1(k)$ and $\phi_2(k)$, where the process
of particle generation is not deterministic but involves some success
probability: let $p_1$ ($p_2$) be the probability that the first (second)
particle is successfully created. In order to keep the discussion simple,
we assume $\phi_1(k)$ and $\phi_2(k)$ to be orthonormalized $\langle
\phi_m | \phi_n \rangle = \delta_{mn}$. The initial state can be written
as a density matrix $\Gamma(\rho)/Z$ with the one-particle density matrix
\begin{equation}\label{eq:two_incoh}
	\rho = \frac{p_1}{1-p_1} |\phi_1\rangle \langle \phi_1 | +
	\frac{p_2}{1-p_2} | \phi_2 \rangle \langle \phi_2 |.
\end{equation}
The weights in (\ref{eq:two_incoh}) are chosen to make the single particle
occupation operator $\eta = \rho/(1+\rho)$ have the form
\begin{equation}\label{eq:number_op}
	\eta= p_1 |
	\phi_1 \rangle \langle \phi_1 | + p_2 |\phi_2 \rangle \langle \phi_2 |,
\end{equation}
i.e., $p_1$ ($p_2$) are the probabilities to occupy the state 1 (2). The
normalized density matrix [we use the fact that $Z = \text{Tr}_{F_a}
\Gamma(\rho) = \det(1+\rho) = 1/(1-p_1)(1-p_2)$]
\begin{multline}\label{eq:dens_matrix}
	\Gamma(\rho)/Z = (1-p_1)(1-p_2) \oplus p_1 (1-p_2)
	|\phi_1 \rangle \langle \phi_1 | \\
	+ p_2 (1-p_1) | \phi_2 \rangle \langle \phi_2|
	\oplus 
	p_1 p_2 |\phi_1
	  \rangle \langle \phi_1| \otimes | \phi_2 \rangle \langle \phi_2|
\end{multline}
consists of three terms: The first term describes the zero particle sector
which occurs with probability $(1-p_1)(1-p_2)$. The second term involves
one particle states: $p_1(1-p_2)$ [$p_2(1-p_1)]$ is the probability that
only the first [second] particle is created. The third term shows that
the probability to observe a Slater determinant of both states $\phi_1$
and $\phi_2$ is $p_1 p_2$; we have omitted terms which involve tensor
products of more than one projector on the same state as they have no
weight on the antisymmetric part of the Hilbert space. The generating
function of full counting statistics is given by (\ref{eq:finite_temp}),
\begin{align}\label{eq:two_incoh_chi}
  \chi(\lambda) &= \det (1- \eta \mathcal{T}^\infty_{Q_\text{R}}+ \eta
  \mathcal{T}^\infty_{Q_\text{R}} e^{i\lambda}) \nonumber\\
	&=
	(1- p_1 T + p_1 Te^{i\lambda}) (1- p_2 T + p_2 T e^{i\lambda})
\end{align}
for the simplest case of asymptotic scattering with an energy-independent
transmission probability, $T_k = T$.

\section{Constant Voltage}

Many results in the literature so far have been obtained
in the stationary regime where a constant voltage $V$ is
applied across the wire for long measuring times $teV/\hbar \gg
1$.\cite{levitov:93,muzykantskii:94,nazarov:02,pilgram:03} Here, we
discuss a wave-packet analog of the constant-voltage case. Contrary to the
discussion in Sec.~\ref{sec:train} involving a nonstationary finite train
of $N$ particles with the spin counter measuring all the time $t\to\infty$,
here, we consider a stationary situation in the thermodynamic limit ($N,L
\to \infty$ with fixed density $n = N/L$, $L$ the system size) with two
reservoirs disbalanced by the applied voltage $V$ and the counting extending
over a finite time $t$.

We start with $N$ particles residing in (left incident) scattering states
\begin{equation}\label{eq:scattering_states}
  \varphi_{k}(x) =\! 
  (e^{i k x} + r_k e^{-i k
  x})\Theta(-x) + t_k e^{i k x} \Theta(x),
\end{equation}
with energies $\hbar\varepsilon=\hbar v_{\F}k$ between $E_{\F}$ and
$E_{\F}+eV$.  The scatterer is positioned at the origin.  In order to
regularize the problem, we go over to wave packets $\phi_m(x)$: We split the
momentum interval $[0,eV/v_{\F}]$ into compartments of width $\hbar \varkappa=
eV/v_{\F}N$ and define the weights
\begin{equation}\label{eq:wf_in}
  f_m(k) = \begin{cases}
    \sqrt{{2\pi}/{\varkappa}}, & \varkappa(m-1) \leq k \leq \varkappa m, \\ 
    0, & \text{elsewhere},
  \end{cases}
\end{equation}
with $m \in \{1,\dots,N\}$. With the real weights $f_m(k)$, the (normalized)
wave packets 
\begin{equation}\label{eq:wf_volt_in}
  \phi_m(x) =
  \int \frac{dk}{2\pi}\, f_m(k) \varphi_{k} (x)
\end{equation}
define states centered around the origin. Note that adding arbitrary
global phases to the wave packets $\phi_m(x)$ does not change their Slater
determinant (up to a trivial global phase of the many-body wave function,
see (\ref{eq:init})).  Keeping $V$ constant and letting $\varkappa \to 0$,
the wave packets spread out in real space, the particle number $N$ goes
to infinity, the homogeneous particle density assumes the finite value
$eV/2\pi\hbar v_{\F}$, and the resulting current $\langle \mathcal{I}
\rangle = (e/h) T V$ is constant in time, cf.\ (\ref{eq:average}). This
procedure then properly emulates the constant voltage setup, as it generates
the identical zero temperature density matrix as the one obtained in a
second quantization formulation by filling scattering states within the
interval of width $eV$.

In making use of the expression (\ref{eq:determinant}), we need the time
evolution of the wave packets,
\begin{equation}\label{eq:wf_volt_out}
  \phi_m(x;t) = \int\!\frac{dk}{2\pi}\,e^{-iv_{\F} k t}
  f_m(k) \varphi_{k}(x),
\end{equation}
as well as the counting operator $\mathcal{Q}_t=\int_I dx |x\rangle \langle
x|$ projecting particles on the space interval $I=[x_0,x_0+v_{\F}t]$, where
we assume the counter to be placed to the right of the origin, $x_0>0$. 

\subsection{Generalized binomial statistics}

The characteristic function $\chi_t(\lambda)$ is the determinant of the
matrix, cf.\ Eq.~(\ref{eq:determinant}),
\begin{align}\label{eq:matrix}
  \langle \phi_m | 
  &e^{i\lambda \mathcal{U}^\dagger \mathcal{Q}_t \mathcal{U}}
  |\phi_n \rangle  
  = \langle \phi_m(t) | e^{i \lambda \mathcal{Q}_t} 
  | \phi_n(t) \rangle \\
  &= \int dx 
  \phi_m(x;t)^* \langle x|e^{i \lambda \mathcal{Q}_t} | x \rangle
  \phi_n(x;t)\nonumber\\
  &=\delta_{mn}\!  + 
  (e^{i\lambda}- 1)\mathsf{Q}_{mn}, \nonumber\\
  \chi_{t} (\lambda) 
  &= \det [\delta_{mn} + (e^{i\lambda} -1 ) \mathsf{Q}_{mn}]
  \label{eq:gbs_V}
\end{align}
with
\begin{equation}\label{eq:matrixQ}
  \mathsf{Q}_{mn} = \int \frac{dk' dk}{4\pi^2} t^{\vphantom *}_k
  f^{\vphantom *}_n(k) K_t(k-k') t^*_{k'} f_m^* (k')
\end{equation}
and the kernel
\begin{align}\label{eq:kernel}
  K_t(q) &= \int_{x_0}^{x_0+v_{\F} t} \!\!\!\!\!\!\!dx\,
  e^{i q (x-v_{\F} t)} = \frac{e^{i q x_0} (1- e^{-i
  q v_{\F} t}) } {i q} \nonumber\\
  &= 2 e^{i q ( x_0- v_{\F} t/2)} \frac{\sin(q v_{\F} t/2)}{q}.
\end{align}
The matrix $\mathsf{Q}_{mn}$ is a Hermitian matrix with real eigenvalues
$0\leq \tau_m(t) \leq 1$, hence the associated full counting statistics
is generalized binomial for all times. The same result can be retrieved
from Ref.\ \onlinecite{abanov:08} with an appropriate choice for the
time-dependent scatterer. In the following, we discuss various limits for
the generating function $\chi_t(\lambda)$.

\subsection{Short measuring time}\label{sec:short}

Assuming that $N$ is large enough so that $t_k$ does not change appreciably
over the interval $\varkappa$, i.e., $\varkappa \partial_k t_k \ll 1$, the
amplitude $t_k$ can be taken out of the integral in Eq.~(\ref{eq:matrixQ}).
Assuming furthermore that the measurement time $t$ is short, $|q| v_{\F}
t \leq t eV/\hbar \ll 1 $, we can expand $K_t(q)$ and obtain (to lowest
order in $teV/\hbar$)
\begin{align}\label{eq:chi_volt_short}
  \chi_{t}^{\ll} (\lambda) &= \det \Bigl[ \delta_{mn} +  (e^{i\lambda}-1)
  t^*_{\varkappa m} t^{\vphantom *}_{\varkappa n} \varkappa v_{\F} t\nonumber\\
  &\qquad\qquad\times e^{i \varkappa x_0 (n-m)} 
  \frac{2\sin^2(\varkappa x_0/2)}{\pi\varkappa^2 x_0^2}
  \Bigr].
\end{align}
The second term involves a matrix product $(v_1, v_2, \dots, v_N)^\dagger
(v_1, v_2, \dots, v_N)$ of a vector and its dual, where $v_m = t_{\varkappa
m}e^{i \varkappa x_0 m}$, and hence can be written as a projector, in
Dirac notation, $\mu |v\rangle \langle v|$ with $\mu = 2 
(e^{i\lambda}-1)\varkappa v_{\F} t \sin^2(\varkappa x_0/2)/\pi\varkappa^2
x_0^2$. The determinant $\det (1 + \mu|v\rangle \langle v|)$ then is
given by the product of eigenvalues $1+\mu\langle v|v\rangle$ (in the
direction of $|v\rangle$) and 1 (in the complement), $\det (1 + \mu |v\rangle
\langle v|) = 1 + \mu \langle v | v \rangle$, and we obtain
\begin{align}\label{eq:chi_volt_short2}
  \chi_{t}^{\ll} (\lambda) &= 1+(e^{i\lambda}-1) \varkappa v_{\F} t \frac{2
  \sin^2 (\varkappa x_0/2)}{\pi \varkappa^2 x_0^2} \sum_{m=1}^N T_{\varkappa m}
  \nonumber\\
  & \stackrel{(\varkappa \to 0)}\longrightarrow 
  1 + \alpha (e^{i \lambda} -1 ),
\end{align}
with (note that $N\varkappa = eV/\hbar v_{\F}$)
\begin{equation}\label{eq:alpha}
    \alpha  = t v_{\F} \int_0^{e V/\hbar v_{\F}} \!\! \frac{dk}{2\pi} T_k 
    \leq t eV / 2\pi \hbar \ll 1
\end{equation}
(note that the dependence on the counter position $x_0$ has disappeared from
the parameter $\alpha \propto t$ in the constant voltage limit $\varkappa \to
0$). Pushing the calculation to higher order in $teV/\hbar$, we find that in
the expansion of $\chi^\ll_t(\lambda)$ both, second- and third-order terms,
vanish and the next correction appears only in fourth order,
\begin{eqnarray}\nonumber
  \Delta \chi^\ll_t(\lambda) \!\! &=&\!\! \frac{(e^{i\lambda} - 1)^2
  (t v_{\F})^4}{24} \!\! \int_0^{eV/\hbar v_{\F}} \!\!
  \frac{dk dk'}{(2\pi)^2} T_k T_{k'} (k - k')^2
  \\ \label{eq:delta_gordey}
  &\leq& \Bigl(\frac{e^{i\lambda}-1}{24 \pi}\Bigr)^2 
  \Bigl(\frac{teV}{\hbar}\Bigr)^4;
\end{eqnarray}
hence, for short measuring times the majority of counts involve either no or
a single particle, while the observation of two-particle events $P_{2}\leq
(teV/\hbar)^4/(24 \pi)^2$ is strongly suppressed, a consequence of the
Pauli exclusion principle. Note that in the short measuring time limit,
the specific nature of the counting device matters. Above, we have assumed
that all intrinsic timescales of the counter are much shorter than the
measuring time.  Furthermore, we have neglected the effect of the Fermi sea
which will produce an additional contribution. Nevertheless, even modeling
the counter more realistically, the Pauli principle with its reduction of
two- and more-particle events is expected to reveal itself. Furthermore,
it is possible to realize an experiment where the effect of the additional
Fermi sea is absent: by applying a voltage to a quantum wire which is larger
than the Fermi energy, particles incident from the right are blocked by the
band bottom and only left going states within an energy interval $E_{\F}$
(replacing the bias $eV$) contribute to the particle current.\cite{brown:89}

Note that the generalized binomial statistics Eq.~(\ref{eq:fcs_2}) reduces
to the simple Poissonian result
\begin{equation}\label{eq:cum_gen}
  \log \chi(\lambda) = \sum_m \tau_m (e^{i\lambda}-1)
\end{equation}
in the limit of small generalized transmission probabilities $\tau_m \ll 1$;
the result then only depends on one parameter $\sum_m \tau_m = \mathrm{tr}
\mathcal{T}_{Q_t}$. In the limit of short measuring times, the smallness
of the transmission eigenvalues is imposed by the small space interval
in the projection $\mathcal{Q}_t$ and $\sum_m \tau_m = \alpha$, see Eq.\
(\ref{eq:alpha}). The same result is obtained in the long time limit, see
(\ref{eq:chi_volt_long2}), provided the transmission probabilities $T_k$
themselves are small.

\subsection{Large measuring times}\label{sec:asym}

In the asymptotic limit of $t \to \infty$, the kernel $K_t(q)$ ensures
energy/momentum conservation, rendering the problem diagonal in the momentum
basis.\cite{levitov:93,schonhammer:07} However, adopting the $t \to \infty$
asymptotic limit is incompatible with a regular derivation of a finite result.
Here, we consider instead the case of large but finite measuring time $t$,
while adopting the limit of infinite particle number $N\to\infty$ when letting
the width $\hbar\varkappa = eV/v_{\F}N$ go to zero at constant voltage $V$.

In the limit $N\to\infty$, the characteristic function $\chi_t(\lambda)$,
which is the determinant of the matrix in Eq.~(\ref{eq:matrix}), cf.\
Eq.~(\ref{eq:determinant}), is given by
\begin{equation}\label{eq:det_asym}
  \chi_t(\lambda) = \det ( 1 -  \mathcal{P} \mathcal{T}
    \mathcal{Q}_t + \mathcal{P} \mathcal{T}
    \mathcal{Q}_t e^{i\lambda})
\end{equation}
where $\mathcal{P}= \int_0^{eV/\hbar v_{\F}} (dk/2\pi) |k\rangle \langle k|$
is the projector on the subspace of occupied states; the form
(\ref{eq:det_asym}) can be obtained from (\ref{eq:matrix}) introducing the
projector $\mathcal{P}$ to extend the determinant over the whole Hilbert
space, cf.\ Eqs.~(\ref{eq:block}) and (\ref{eq:determinant_general}), and
using the determinant identity $\det(1+\mathcal{A}\mathcal{B})=
\det(1+\mathcal{B} \mathcal{A})$ to shuffle $t_k'^*$ to the left of $K_t$,
which itself is the momentum representation of the projector $\mathcal{Q}_t$.
The expression Eq.~(\ref{eq:det_asym}) then corresponds to
Eq.~(\ref{eq:determinant_general}) with the substitution $\mathcal{T}_Q =
\mathcal{T} \mathcal{Q}_t$. As $\mathcal{Q}_t$ is a projector,
$\mathcal{Q}_t^2 = \mathcal{Q}_t$, we can rewrite Eq.~(\ref{eq:det_asym}) as
$\det [1+ (e^{i\lambda}-1) \mathcal{Q}_t \mathcal{P} \mathcal{T}
\mathcal{Q}_t]$. This determinant only needs to be calculated in the subspace
$\mathcal{Q}_t H$ as the matrix is unity in the complement. In the subspace
$\mathcal{Q}_t H$, we use the orthonormal real-space (rather than $k$-space,
see Eq.~(\ref{eq:wf_in})) basis
\begin{equation}\label{eq:trunc}
  g_l(x) = \begin{cases}
    1/\sqrt{\epsilon}, & \epsilon(l-1) \leq x-x_0 \leq \epsilon
    l, \\ 
    0, & \text{elsewhere},
  \end{cases}
\end{equation}
with $\epsilon = t v_{F} / L$ the width of a real space segment and $l\in
\{1,\dots,L\}$. The matrix elements of $\mathcal{P} \mathcal{T} $ assume the
form
\begin{align}\label{eq:m_elem}
  \langle g_l | \mathcal{P} \mathcal{T} | g_m \rangle &=
  \int_0^{eV/\hbar v_{\F}} \frac{dk}{2\pi} T_k \langle g_l | k\rangle
  \langle k | g_m \rangle \\
  &=\int_0^{eV/\hbar v_{\F}} \!\! \frac{dk}{2\pi} T_k \frac{4 \sin^2 (\epsilon
  k/2)}{\epsilon k^2} e^{i (l-m) k \epsilon}, \nonumber 
\end{align}
i.e., they form a Toeplitz matrix. Applying Szeg\H{o}'s theorem and taking the
limit of large $t$ and $L$ with $\epsilon = v_{\F} t /L$ fixed but small
($\epsilon \ll \hbar v_{\F}/eV$, hence $4 \sin^2 (\epsilon k/2)/\epsilon
k^2 \approx \epsilon$) we obtain the generating function
\begin{equation}\label{eq:chi_volt_long2}
  \log \chi_t (\lambda) = t v_{\F} \!\! \int_0^{eV/\hbar v_{\F}}
  \frac{dk}{2\pi} \log ( 1- T_k + T_ k e^{i \lambda} ),
\end{equation}
cf., App.~\ref{sec:szego}. 

Using the generalization of Szeg\H{o}'s theorem (Fisher-Hartwig
conjecture, see (\ref{eq:szego_det}) and (\ref{eq:log})) \cite{basor:79},
it is possible to calculate the next order term. As the argument of the
logarithm (cf.~(\ref{eq:log}), note that $x(\theta)$ is to be replaced by
$\sum_{n\in\mathbb{Z}} P_{(\theta+2\pi n)/\epsilon} [1 + (e^{i\lambda}
-1)T_{(\theta+2\pi n)/\epsilon} ]$ with $P_k = 1$, $k\in[0,eV/\hbar
v_{\F}]$ and $P_k=0$ otherwise) exhibits discontinuities at $k=0$ and
$k=eV/\hbar v_{\F}$, the correction to the leading term is given by the
two contributions originating from the jumps at $k = 0, eV/\hbar v_{\F}$,
\begin{equation}\label{eq:delta_chi}
  \Delta \log \chi_t(\lambda) = \frac{ \log(t/t_0) }{4\pi^2} \sum_{
  \makebox[12pt]{$\scriptstyle k= 0,eV/\hbar v_{\F}$}}
  \log^2[ 1+ (e^{i\lambda} -1 ) T_k ], 
\end{equation}
with $t_0$ some small time cutoff; this result leads to logarithmic corrections
for the second-order and all higher cumulants. For the noise, the correction
is given by\cite{muzykantskii:03,schonhammer:07}
\begin{equation}\label{eq:delta_n2}
  \Delta  \langle\langle n^2 \rangle \rangle = \frac{T_0^2 + T_{eV/\hbar
  v_{\F}}^2}{2\pi^2}
  \log (t/t_0).
\end{equation}
Here, the logarithmic corrections in Eqs.~(\ref{eq:delta_chi}) and
(\ref{eq:delta_n2}) are due to fluctuations in the number of particles
in a finite interval of length $v_{\F} t$. Therefore, fluctuations
do not disappear for $T=1$ as in Eqs.~(\ref{eq:delta_chi_N}) and
(\ref{eq:delta_n2_N}) where the number of particles is fixed and noise stems
only from partitioning. In addition to the noise originating from the voltage
bias, there is an equilibrium contribution due to the Fermi sea at any
finite measuring time, cf.\ Sec.~\ref{sec:short}. For asymptotically large
times, the first contribution grows logarithmically in time.\cite{levitov:96}

The reason why Szeg\H{o}'s theorem is applicable to the matrix
Eq.~(\ref{eq:m_elem}) is the presence of time translation invariance:
matrix elements between states localized at two different places/times
depend only on their space/time separation and hence they form a Toeplitz
matrix. The same reasoning does not apply to the momentum basis and that
is why we could not apply Szeg\H{o}'s theorem directly to the matrix in
Eq.~(\ref{eq:matrix}).  The result Eq.~(\ref{eq:chi_volt_long2}) (and its
generalization to finite temperatures, cf.\ Eq.~(\ref{eq:finite_temp_2})) has
been found by Sch\"onhammer\cite{schonhammer:07} using a double projection
in his counting procedure: instead of relying on Szeg\H{o}'s theorem,
use has been made of the relation $\log \det(1+M) = \text{tr} \log(1+
M)$, followed by an expansion of the logarithm. Evaluating the trace of
each term, the phase factors (see Eq.~(\ref{eq:kernel})) appearing in the
cyclic product of the kernel $K_t$ cancel mutually.  In the long-time limit,
the kernels become diagonal (ensuring energy-conservation) with one of them
contributing a factor of $t$, thus rendering the cumulant generating function
$\log\chi_t(\lambda)$ linear in $t$. In alternative approaches, use has been
made of a mapping onto the Riemann-Hilbert problem \cite{muzykantskii:03}
(this procedure enables the calculation of the leading term as well as the
logarithmic correction Eq.~(\ref{eq:delta_chi})), or of time periodicity
\cite{levitov:96,ivanov:97} introduced in order to render $\log \chi_t$
extensive in $t$ (this way, only the term linear in $t$ is obtained).

\subsection{Fano factor for intermediate regime}

\begin{figure*}[!t]
  \centering
  \includegraphics{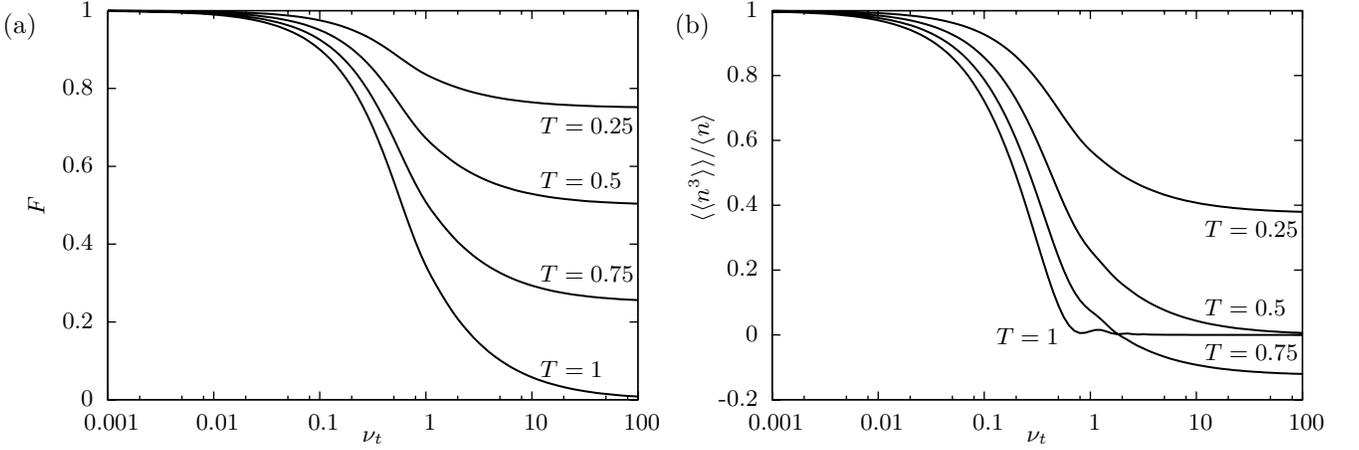}
  \caption{%
  Fano factor $F=\langle\langle n^2\rangle\rangle/\langle n \rangle$ in (a)
  and third moment $\langle\langle n^3 \rangle\rangle/\langle n \rangle$
  in (b) for constant voltage with energy independent transmission
  probabilities $T=0.25,~0.5,~0.75,~1$, as a function of the incident
  particle number $\nu_t = t e V/ 2\pi \hbar$. Note that the Fano factor
  approaches the binomial value $F=1-T$ for $\nu_t\gg 1$ whereas for
  $\nu_t \ll 1$ it is always close to 1 irrespective of the transmission
  probability $T$. The third cumulant interpolates as a function of $\nu_t$
  between the Poissonian value $\langle\langle n^3 \rangle\rangle/\langle
  n \rangle= 1$, cf.\ Eq.\ (\ref{eq:cum_gen}), and the binomial result
  $\langle\langle n^3 \rangle\rangle/\langle n \rangle= T (1-T)(1-2T)$,
  cf.\ Eq.~(\ref{eq:chi_volt_long2}).  The oscillations (especially around
  $\nu_t \approx 1$ for $T=1$) are due to the sharp edge in the occupation
  number at $k = eV/\hbar v_{\F}$.
  }\label{fig:fano}
\end{figure*}

In order to understand the crossover between the short and long time
behavior of the carrier distribution, we calculate the Fano factor $F$ and
present the result as a function of $\nu_t = teV/2\pi\hbar$ (the incident
particle number during time $t$) in Fig.~\ref{fig:fano}(a) for several
values of the transmission coefficient $T$ (for a scatterer with energy
independent transmission). For small times, the distribution is Poissonian
and hence $F(\nu_t\to 0) \to 1$. The binomial distribution valid at large
times provides the asymptotics $F(\nu_t\to\infty) \to (1-T)$. In order
to find the crossover in between, we determine the matrix $\mathsf{Q}$
(see Eq.~(\ref{eq:matrixQ})),
\[
  \mathsf{Q}_{mn}
  \stackrel{(\varkappa \to 0)}\longrightarrow
  t_{\varkappa m}^* t_{\varkappa n}^{\vphantom *} e^{i (n-m) 
  \varkappa (x_0 - v_{\F} t)} 
  \frac{\sin[ (n-m) \varkappa v_{\F} t/2]}{\pi(n-m)},
\]
in terms of which the characteristic function assumes the simple form
(\ref{eq:gbs_V}) and hence $\log \chi_{t} (\lambda) = \text{tr}\log[
\delta_{mn} + (e^{i\lambda} -1 ) \mathsf{Q}_{mn}]$ (again, we consider
the limit $\varkappa\to0$ at fixed voltage $V$).  The average transmitted
charge $\langle n \rangle= -i\partial_\lambda \log \chi_{t}|_{\lambda=0}$,
\begin{equation}\label{eq:average}
  \langle n \rangle = \text{tr} \, \mathsf{Q} = t v_{\F}  \int_0^{e V/\hbar
  v_{\F}} \!\! \frac{dk}{2\pi} T_k,
\end{equation}
grows linearly with the measuring time $t$; the above result coincides
with those obtained from the short and long time expressions
(\ref{eq:chi_volt_short2}) and (\ref{eq:chi_volt_long2}). The
noise $\langle\langle n^2 \rangle\rangle= -\partial^2_\lambda
\log\chi_{t}|_{\lambda=0}$ assumes the form
\begin{align}\label{eq:noise}
  \langle\langle n^2 \rangle \rangle &= \text{tr} \, \mathsf{Q} - \text{tr}
  \, \mathsf{Q^2} \\
  &= \langle n \rangle - \int_0^{e V/\hbar
  v_{\F}} \!\! \frac{dk' dk}{\pi^2} T_{k'}
  T_k \frac{\sin^2 [ (k'-k) v_{\F} t/2]}{(k'-k)^2}. \nonumber
\end{align}
(in the limit $\varkappa \to 0$ considered here, both momenta do not
depend on the position $x_0$ of the counter, as the wave packets are
infinitely spread).  In order to keep the analysis simple, we assume an
energy independent transmission probability, $T_k = T$, over the interval
$[0,e V/\hbar v_{\F}]$.  The average charge then is given by
\begin{equation}\label{eq:average2}
  \langle n \rangle = T t e V / 2\pi \hbar =T \nu_t.
\end{equation}

The Fano factor $F=\langle\langle n^2 \rangle\rangle / \langle n \rangle$
can be cast into the form
\begin{equation}\label{eq:fano_factor2}
  F =  1 - T f(\nu_t) 
\end{equation}
with 
\begin{equation}\label{eq:f}
  f(\nu_t) = \int_{-1}^1 dx (1- |x|) \frac{\sin^2(\pi \nu_t x)}{\pi^2 \nu_t x^2}.
\end{equation}
For small times $\nu_t \ll 1$,
\begin{equation}\label{eq:f_small}
  f(\nu_t) = \nu_t - \frac{\pi^2}{18} \nu_t^3 + \mathcal{O}(\nu_t^5),
\end{equation}
while $f$ approaches unity in the long time limit $\nu_t\gg1$ ($\gamma
\approx0.5772$ is Euler's constant),
\begin{equation}\label{eq:f_long}
  f(\nu_t) = 1  - \frac{ \log (2\pi\nu_t) + 1 + \gamma}{\pi^2 \nu_t} +
  \mathcal{O}(\nu_t^{-3}).
\end{equation}
The corrections to the simple binomial result produce a logarithmic in
time increase of the noise $\langle\langle n^2 \rangle\rangle$; the result
(\ref{eq:f_long}) coincides with Eq.~(\ref{eq:delta_n2}) for the case
of energy independent scattering probabilities $T_k = T$. This logarithmic
dependence in the noise is due to the fluctuations in the number of
electrons in a finite segment of the wire.\cite{levitov:93} 

Analogously, the third cumulant $\langle\langle n^3 \rangle\rangle$ can
be calculated; the (numerical) results, shown in Fig.~\ref{fig:fano}(b),
interpolate between the Poissonian value $\langle\langle n^3 \rangle\rangle/
\langle n\rangle = 1$ for short times and the binomial result $\langle\langle
n^3 \rangle\rangle/ \langle n\rangle = T(1-T)(1-2T)$ for long measuring
times.

\subsection{Finite temperature}

We consider the case where particles are emitted from a lead at finite
temperature into vacuum, i.e., we assume a single Fermi reservoir of particles
(incident from the left) which are scattered with energy-dependent
transmission probabilities (to the right). At finite temperature, scattering
states are occupied according to the Fermi-Dirac occupation as described by
the one-particle operator $\eta = [e^{\beta (\mathcal{H}-\mu)}+1]^{-1}$. The
characteristic function $\chi_t(\lambda)$ is given by
Eq.~(\ref{eq:finite_temp}), with $\mathcal{T}_Q = \mathcal{T} \mathcal{Q}_t$
and the interval $I = [x_0, x_0 + v_{\F} t]$ defining the projector
$\mathcal{Q}_t$, cf.\ Sec.~\ref{sec:asym}. Following essentially the
calculation in Sec.~\ref{sec:asym}, i.e., calculating the determinant in the
basis (\ref{eq:trunc}) and applying Szeg\H{o}'s theorem, the result
\begin{equation}\label{eq:chi_finite_temp}
  \log \chi_t(\lambda) = t v_{\F} \int \frac{dk}{2\pi} \log \Bigl[1+
  T_k n_\text{L}(k) (e^{i\lambda}-1)
	\Bigr],
\end{equation}
with $n_\text{L} (k) = \langle k | \eta | k \rangle = [e^{\beta (\hbar
v_{\F}k-\mu)}+1]^{-1}$ can be obtained for fixed but long measurement 
times $t$. For high temperatures at constant particle density [$\beta
\to 0$, $n_\text{L}(k) \approx e^{-\beta (\hbar v_\text{F} k -\mu)}$],
all transmission eigenvalues $\tau_k = T_k / (e^{\beta( \hbar v_{\F}
k -\mu)}+1) \approx T_k e^{-\beta( \hbar v_{\F} k - \mu)}$ approach
zero. The logarithm in Eq.~(\ref{eq:chi_finite_temp}) can be expanded
and the emission statistics for electrons leaving a Fermi reservoir in the
high temperature regime is given by a Poissonian statistics\cite{schottky:18}
\begin{equation}\label{eq:poisson}
  \log \chi_t(\lambda) =  (e^{i\lambda}-1)t v_{\F}  \int \frac{dk}{2\pi} 
    T_k e^{-\beta( \hbar v_{\F} k -\mu)} 
\end{equation}
The Fano factor assumes the value $F= 1$, independent of $T_k$.  

To complete the analysis, we discuss the extension of the constant-voltage
result with two reservoirs to finite temperatures.  We model the setup by two
Fermi reservoirs with occupation numbers $n_{\text{L}/\text{R}}(k)=[\exp[\beta
(\hbar v_{\F}k-\mu_{\text{L}/\text{R}})]+1]^{-1}$ for particles incoming from
the left (L) or right (R), respectively. The voltage enters via the bias of
the chemical potentials $e V = \mu_\text{L} - \mu_\text{R}$. 

Unfortunately, it is not possible to define a projection operator
$\mathcal{Q}$ which acts {\it after} the evolution and which can
emulate the action of the spin counter, cf.\ the discussion below
Eq.~(\ref{eq:wave_pm}). The reason is that there is no way to tell for a
particle outgoing to the left at time $t\to\infty$ whether it was coming in
from the left and was reflected at the scatterer (hence, no counting is done
with the counter to the right of the scatterer) or whether it was coming in
from the right and has been transmitted through the scatterer (hence passing
the spin counter once). One solution to this problem is to perform a first
projective measurement at the initial time; \cite{muzykantskii:03,avron:08}
this corresponds to replacing Eq.~(\ref{eq:finite_temp}) by the expression
\begin{equation}\label{eq:project}
	\chi_t (\lambda) = 
	\frac{\det ( 1 + \rho \, \mathcal{U}^\dagger e^{i\lambda
	\mathcal{Q}_t} \mathcal{U} e^{-i\lambda\mathcal{Q}_t})}{\det(1+\rho)},
\end{equation}
with the occupation-operator $\eta = \rho/(1+\rho) = \int (dk/2\pi)
|k\rangle_\text{in} \, \text{diag} [n_\text{L}(k),n_\text{R}(k)]\,
{}_\text{in}\langle k |$, the single-particle evolution $\mathcal{U}$
involving the scatterer but not the spin counter, cf.\ Eq.~(\ref{eq:prop}),
and $\mathcal{Q}_t$ a projector emulating the counting measurement of
transmitted particles via projection of the wave functions onto the lead to
the right of the scatterer.  The additional factor $\exp(-i \lambda
\mathcal{Q}_t)$, as compared to (\ref{eq:finite_temp}), corresponds to the
additional measurement before the evolution. 

\begin{figure}[t]
	\centering
	\includegraphics{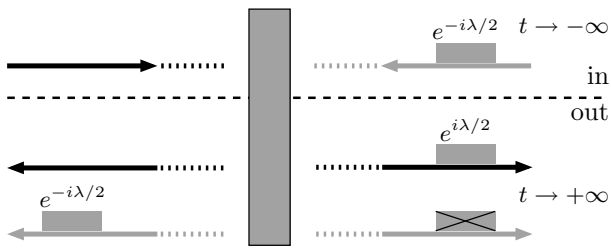}
	\caption{%
	Sketch of incoming ($t\to-\infty$, above the dashed line) and
	scattered (large gray box) outgoing ($t\to+\infty$, below the
	dashed line) states measured by a spin counter (small gray boxes)
	placed to the right of the scatterer. Left-incoming and scattered
	states are described by black arrows, right-incoming and scattered
	states correspond to grey arrows. The action of the spin counter
	(in the up state) is included in the expression $\mathcal{U}_{+}$,
	see Eq.~(\ref{eq:spin_evolution}): The evolution $\mathcal{U}_{+}$
	involves two projections, one described by $\exp(-i \lambda
	\mathcal{Q}_t/2)$ at time $t\to-\infty$ and the second $\exp(+i
	\lambda \mathcal{Q}_t/2)$ at time $t\to\infty$. The right (grey
	arrow) incoming state at $t\to-\infty$ acquires a phase factor
	$\exp(-i\lambda/2)$ (gray box on the right) in the first projection;
	this provides the correct counting field for its transmitted part
	(gray box on the left).  The phase factor of the reflected part
	is canceled by the second counting operator (crossed box on the
	right). The left incoming state is unaffected by the first counting
	and its transmitted part acquires a phase $\exp(i\lambda/2)$
	at time $t\to\infty$.
	}\label{fig:counter}
\end{figure}

In this article, we want to stick with the spin counter as a measurement
apparatus. Contrary to the situation in Sec.~\ref{sec:generalizations},
the action of the spin counter cannot be modeled by a projection
onto the outgoing states, i.e., the operators $\mathcal{U}_{\pm}$
do not separate any more into factors describing the scatterer and
the counter, $\mathcal{U}_{\pm} \neq e^{\pm i \lambda \mathcal{Q}_t/2}
\mathcal{U}$.  Therefore, we have to make use of the full evolution operators
$\mathcal{U}_{\pm}$, $|\Psi^\pm_\text{out}\rangle = \Gamma( \mathcal{U}_\pm)
| \Psi \rangle$, in the presence of both the scatterer and the spin counter,
where the index `$\pm$' refers to the two spin states of the counter.
The overall evolution (cf.\ Fig.~\ref{fig:counter}) then can be written as
\begin{equation}\label{eq:spin_evolution}
	\mathcal{U}_\pm = e^{\pm i \lambda \mathcal{Q}_t/2} \,
	\mathcal{U}  e^{\mp i \lambda \mathcal{Q}_t/2},
\end{equation}
where $\mathcal{U}$, cf.\ Eq.~(\ref{eq:prop}), is the evolution without
accounting for the presence of the spin-counter.
The generating function for the full counting statistics assumes the form
\begin{equation}\label{eq:spin_counter}
	\chi_t (\lambda) = \frac{\det ( 1 + \rho \, 
	 e^{-i \lambda \mathcal{Q}_t/2}
	  \mathcal{U}^\dagger e^{i\lambda
	  \mathcal{Q}_t} \mathcal{U} 
		e^{-i \lambda \mathcal{Q}_t/2} ) }{ \det (1+\rho)}.
\end{equation}
The two counting procedures Eqs.~(\ref{eq:project}) and
(\ref{eq:spin_counter}) agree if the particles are only incident
from the left, as the additional counting factors, compared to
Eq.~(\ref{eq:finite_temp}), contribute unity. For particles incoming
from both left and right the two counting procedures do not necessarily
coincide; only if $\mathcal{Q}_t$ commutes with $\rho$, we can shift the
factor $e^{-i \lambda \mathcal{Q}_t/2}$ to the left of $\rho$ and then
cyclically permute the factors in the second term of the determinant to
assert the equivalence of (\ref{eq:project}) and (\ref{eq:spin_counter}).

The interpretation of Eq.~(\ref{eq:spin_counter}) as the generating function
for the full counting statistics using the spin counting procedure faces
problems, since Eq.~(\ref{eq:spin_evolution}) is not necessarily $2\pi$
periodic and hence the counting may involve a noninteger number of particles.
In certain situations, however, the spin counter nevertheless leads to
sensible results. In particular, for asymptotically long measuring times
$t\to\infty$, the counting projection operator $\mathcal{Q}_t$ becomes
basically diagonal in the energy or momentum basis. Commuting $\exp(-i\lambda
\mathcal{Q}_t/2)$ with $\rho$ and repeating the calculation in
Sec.~\ref{sec:asym}, i.e., calculating the determinant in the basis
(\ref{eq:trunc}) and applying Szeg\H{o}'s theorem, the result
\begin{align}\label{eq:finite_temp_2}
	\log \chi_t^{\!\gg} \!(\lambda) =& 
	t v_{\F} \!\! \int \!\frac{dk}{2\pi} \log \Bigl[1+
  T_k \bigl\{n_\text{L}(k) [1-n_\text{R}(k)] (e^{i\lambda} \!-\! 1) \nonumber\\
  & \quad+
  n_\text{R}(k) [1-n_\text{L}(k)] (e^{-i\lambda}-1) \bigr\}
      \Bigr],
\end{align}
is obtained; alternatively, the result (\ref{eq:finite_temp_2}) can be
obtained by expanding the determinant using the relation $\log \det (1+M) =
\sum_{k=1}^\infty (-1)^k \text{tr} M^k /k$ and determining the leading
contribution in each order.\cite{schonhammer:07} Away from the asymptotic
limit (including also the calculation of next-to-leading order corrections)
the above commutation cannot be carried out and half-integer charges might
show up.\cite{shelankov:03} A thorough discussion of the equivalence of the
counting procedures (\ref{eq:project}) and (\ref{eq:spin_counter}) for finite
measuring times is still lacking.

\section{Conclusion}\label{sec:conc}

We have used the first-quantized wave packet formalism to calculate the
generating function $\chi_N(\lambda)$ of full counting statistics of
fermionic particles in various physical situations, such as $N$ particles
incident in Slater determinant states of rank 1 (nonentangled), rank 2
(entangled), or incoherent superpositions of Slater determinants in Fock
space with undetermined particle number. Our formalism captures various
features such as energy dependent scattering probabilities as well as
time-dependent scattering and time-dependent counting. 

We have presented our results in determinantal form, with further
simplifications explicitly unveiling a generalized binomial statistics in
various cases. Applications of our results include a classification of
possible statistical behavior of two-particle scattering events and a
particularly simple singlet-triplet and entanglement detector.  In the context
of coherent transport of noninteracting (degenerate) fermions, the natural
reference point in the discussion of statistical properties is the binomial
distribution; energy dependent scattering naturally shifts the noise into the
sub-binomial (or generalized binomial) regime, whereas additional correlations
through entanglement can generate super-binomial noise statistics. 

Our results, calculated at zero temperature, remain valid for $\beta^{-1}\ll
\hbar v_{\F}/\xi$, i.e., sufficiently narrow wave packets with a small width
$\xi$ in real space.  Furthermore, we have calculated the generating function
for the constant voltage case in the long-time limit for any temperature.  For
short measuring times our results are valid in the temperature regime
$\beta^{-1} \ll eV$ and we have found a strong suppression of $P_{n \geq 2}$
due to Pauli blocking.

The central element underlying the appearance of a (sub-)binomial statistics
in fermionic systems is the absence of interparticle interactions and
entanglement.  This result remains valid for a time-dependent scattering
potential and finite temperature. We have analyzed the modification
introduced by entanglement and have found that super-binomial statistics
may be generated. The inclusion of interaction, particularly within
the scatterer where interacting particles become entangled, remains an
interesting open problem.

\acknowledgments

We acknowledge fruitful discussions with Andrei Lebedev and Dima Ivanov
and financial support from the CTS-ETHZ, the Swiss National Foundation,
the Russian Foundation for Basic Research (08-02-00767-a), and the program
`Quantum Macrophysics' of the RAS.

\appendix

\section{(Strong) Szeg\H{o} Theorem}\label{sec:szego}

The (strong) Szeg\H{o} theorem applies to Toeplitz matrices and reduces the
calculation of the asymptotic behavior of their determinants to a simple
integration (plus summation) problem. We define a Toeplitz matrix starting
from a complex-valued periodic function $a(\theta)$ with $a(\theta+2\pi)
=a(\theta)$. In addition, we require that its winding number with respect
to the origin is equal to zero. We define the Fourier coefficients
\begin{equation}\label{eq:fourier}
  a_m = \int_0^{2\pi} \frac{d\theta}{2\pi} a(\theta) e^{-im\theta}
\end{equation}
and the associated $N \times N$ Toeplitz matrix with elements
\begin{equation}\label{eq:toeplitz} 
  [\mathsf{A}_N (a)]_{m,n} = a_{m-n}
\end{equation}
depending only on the difference between the indices $m$ and $n$ (banded
matrix), $m,n = 1,\dots,N$. The strong form of the Szeg\H{o} theorem
\cite{szego:52,simon:05} states that 
\begin{align}\label{eq:szego2}
  &\log \det \mathsf{A}_N (a) \\
  &\qquad\quad  
  \sim N \, [\log a]_0 + \sum_{n=1}^\infty n [\log a]_n [\log a]_{-n}
  \nonumber
\end{align}
asymptotically for $N\to\infty$, with
\begin{equation}\label{eq:loga}
  [\log a]_n = \int_0^{2\pi} \frac{d\theta}{2\pi} \log [a(\theta)] e^{-i n \theta}
\end{equation}
the Fourier coefficients of $\log[a(\theta)]$. The first term in
(\ref{eq:szego2}) scaling with $N^1$ is the result of Szeg\H{o}'s theorem,
while its strong form applies once the sum in the second term converges ---
this correction then scales with $N^0$.

Given the Toeplitz matrix $\mathsf{X}^f = \mathsf{S}^f+(e^{i\lambda}
-1)\mathsf{T}^f$, cf., Eq.~(\ref{eq:overlap}), we show how to find its
determinant Eq.~(\ref{eq:szego}) in the asymptotic limit of large $N$.
Specifying the matrix elements 
\begin{equation}\label{eq:aelement}
  x_{m-n} = \int \frac{dk}{2\pi}\, |f(k)|^2
  (1- T_k + T_k e^{i\lambda}) e^{i (m-n)k a},
\end{equation}
we find the original periodic function $x(\theta)$ by calculating the Fourier
series
\begin{align}\nonumber
  x(\theta) &= \sum_{m\in \mathbb{Z}} x_m e^{i m \theta}\\
  \label{eq:atheta}
  &= \frac{1}{a} \sum_{m \in \mathbb{Z}}  |f[(\theta+2\pi
  m)/a]|^2 \\
  &\qquad\qquad\times [1- T_{(\theta
  + 2 \pi m)/a} + T_{(\theta + 2 \pi m)/a} e^{i \lambda} ].\nonumber
\end{align}
Note that, while the original function $x(k) = |f(k)|^2 (1- T_k + T_k
e^{i\lambda})$ was defined on the real axis, the new expression $\kappa (k)
= x(\theta = ak)$ is restricted to the first Brillouin zone $k \in
[0,2\pi/a]$.  Fourier transforming the logarithm of $x(\theta)$ according to
Eq.~(\ref{eq:loga}), we obtain the asymptotic expression for the determinant
\begin{align}\label{eq:szego_det}
  \log \det \mathsf{X}^f_N &= N \int_0^{2\pi}\frac{d\theta}{2\pi}
  \,\log[x(\theta)] \nonumber \\
  &\qquad + \sum_{n=1}^\infty n [\log x]_n [\log x]_{-n} + o(1),
\end{align}
consisting of a main term $\propto N$, a first correction staying constant as
$N\to\infty$, and a remaining correction $o(1)$ vanishing as $N \to \infty$.
The (logarithm of the) determinant $\mathsf{S}^f$ in Eq.~(\ref{eq:szego})
is derived by setting $T\equiv 0$ in (\ref{eq:atheta}).  Finally, we
obtain the (log of the) characteristic function by simple subtraction (we
replace the angle $\theta$ on the unit circle $[0,2\pi]$ by $k = \theta/a$
in the first Brillouin zone $[0,2\pi/a]$), to leading order in $N$
\begin{equation}\label{eq:char_app}
  \log \chi_{N}
  (\lambda) = N a \int_0^{2\pi/a} \! \frac{dk}{2\pi} \log(
  1 - \tau_k + \tau_k e^{i \lambda})
\end{equation}
with the effective scattering probabilities
\begin{equation}\label{eq:eff_scatter_app}
  \tau_k = \frac{ \sum_{m \in \mathbb{Z}} | f(k + 2 \pi m/a) |^2 T_{k+2
  \pi m/a} } {\sum_{m \in \mathbb{Z}} | f(k + 2\pi m/a)|^2}.
\end{equation}

For a function $x(\theta)$ which is continuous on the unit circle, i.e.,
$x(2\pi)=x(0)$, the sum in (\ref{eq:szego_det}) converges and the corrections
to (\ref{eq:char_app}) are constant when $N\to \infty$ (and similar for
$s(\theta) = \sum_{m\in\mathbb{Z}} |f[(\theta + 2\pi m)/a]|^2/a$ in the
calculation of $\log \det \mathsf{S}^f_N$). A more subtle situation appears
in the situation where $x(\theta)$ and/or $s(\theta)$ are discontinuous
across the Brillouin zone (Fisher-Hartwig conjecture).\cite{basor:79} This
situation is the usual case as the wave function $f(k)$ is discontinuous at
the Fermi level $k=0$. Thus, the sum in (\ref{eq:szego_det}) is divergent and
the next term in the expansion of (\ref{eq:char_app}) scales with $\log N$
(cf.\ also Eq.~(\ref{eq:delta_chi})),
\begin{multline}\label{eq:log}
  \Delta \log \chi_N (\lambda) =  \\
  \frac{\log^2 [x(0^+)/x(0^-)] - \log^2 [s(0^+)/s(0^-)]}{4 \pi^2}\log N,
\end{multline}
and is followed by a constant term. Here, the number of particles
$N$ is fixed and noise is due to partitioning; hence, the logarithmic
corrections (\ref{eq:log}) have to be attributed to partitioning (and not to
fluctuations in the number of particles as for the constant voltage result
Eq.~(\ref{eq:delta_chi})). This is also consistent with the vanishing of the
correction (\ref{eq:log}) for $T=1$ where $x(\theta)= e^{i\lambda}s(\theta)$
and $x(0^+)/x(0^-)=s(0^+)/s(0^-)$.


\begin{thebibliography}{10}

\bibitem{levitov:93}
L.S.\ Levitov and G.B.\ Lesovik,
 JETP Lett.\ {\bf 58}, 230 (1993).

\bibitem{levitov:96}
L.S.\ Levitov, H.W.\ Lee, and G.B.\ Lesovik,
 J.\ Math.\ Phys.\ {\bf 37}, 4845 (1996).

\bibitem{nazarov}
Yu.V.\ Nazarov, ed.,
 {\em Quantum Noise in Mesoscopic Physics}
 (Kluwer Academic Publishers, 2003).

\bibitem{lesovik:06}
G.B.\ Lesovik, F.\ Hassler, and G.\ Blatter,
 Phys.\ Rev.\ Lett.\ {\bf 96}, 106801 (2006).

\bibitem{peres:84}
A.\ Peres,
 Phys.\ Rev.\ A {\bf 30}, 1610 (1984).

\bibitem{martin:92}
Th.\ Martin and R.\ Landauer,
 Phys.\ Rev.\ B {\bf 45}, 1742 (1992).

\bibitem{lee:95}
H.\ Lee and L.S.\ Levitov,
 cond-mat/9507011 (1995).

\bibitem{ivanov:97}
D.A.\ Ivanov, H.\ Lee, and L.S.\ Levitov,
 Phys.\ Rev.\ B {\bf 56}, 6839 (1997).

\bibitem{lebedev:05}
A.V.\ Lebedev, G.B.\ Lesovik, and G.\ Blatter,
 Phys.\ Rev.\ B {\bf 72}, 245314 (2005).

\bibitem{keeling:06}
J.\ Keeling, I.\ Klich, and L.S.\ Levitov,
 Phys.\ Rev.\ Lett.\ {\bf 97}, 116403 (2006).

\bibitem{feve:07}
G.\ F\`eve, A.\ Mah\'e, J.-M.\ Berroir, T.\ Kontos, B.\ Pla\c{c}ais,
  D.C.\ Glattli, A.\ Cavanna, B.\ Etienne, and Y.\ Jin,
 Science {\bf 316}, 1169 (2007).

\bibitem{hassler:07}
F.\ Hassler, G.B.\ Lesovik, and G.\ Blatter,
 Phys.\ Rev.\ Lett.\ {\bf 99}, 076804 (2007).

\bibitem{entangled}
We use the term `entangled-state' for indistinguishable particles in the sense
  of J.\ Schliemann, J.I.\ Cirac, M.\ Ku\'s, M.\ Lewenstein, and D.\ Loss,
  Physical Review A {\bf 64}, 022303 (2001).

\bibitem{taddei:02}
F.\ Taddei and R.\ Fazio,
 Phys.\ Rev.\ B {\bf 65}, 075317 (2002).

\bibitem{glauber}
K.E.\ Cahill and R.J.\ Glauber, Phys.\ Rev.\ A {\bf 59}, 1538 (1999). Cahill
  \emph{et al.} denote their generating function (generating counting
  probabilities upon taking derivatives) by $\mathcal{Q}(\lambda)$ where their
  $\lambda$ is equivalent to $1-e^{i\lambda}$ in our article.

\bibitem{glauber:65}
R.J.\ Glauber,
 in {\em Quantum Optics and Electronics}, edited by C.\ DeWitt, A.\ Blandin,
  and C.\ Cohen-Tannoudji
 (Gordon and Breach, New York, 1965).

\bibitem{braungardt:08}
S.\ Braungardt, A.\ Sen(De), U.\ Sen, R.J.\ Glauber, and M.\ Lewenstein,
 arXiv:0802.4276 (2008).

\bibitem{burkard:00}
G.\ Burkard, D.\ Loss, and E.V.\ Sukhorukov,
 Phys.\ Rev.\ B {\bf 61}, R16303 (2000).

\bibitem{schonhammer:07}
K.\ Sch\"onhammer,
 Phys.\ Rev.\ B {\bf 75}, 205329 (2007).

\bibitem{szego:15}
G.\ Szeg\H{o},
 Math.\ Ann.\ {\bf 76}, 490 (1915).

\bibitem{szego:52}
G.\ Szeg\H{o},
 Comm.\ S\'em.\ Math.\ Univ.\ Lund pp.\ 228--238 (1952).

\bibitem{levitov:92}
L.S.\ Levitov and G.B.\ Lesovik,
 JETP Lett.\ {\bf 55}, 555 (1992).

\bibitem{lesovik:03}
G.B.\ Lesovik and N.M.\ Chtchelkatchev,
 JETP Lett.\ {\bf 77}, 393 (2003).

\bibitem{levitov:94}
L.S.\ Levitov and G.B.\ Lesovik,
 cond-mat/9401004 (1994).

\bibitem{neder:07}
I.\ Neder and F.\ Marquardt,
 New J.\ Phys.\ {\bf 9}, 112 (2007).

\bibitem{averin:05}
D.V.\ Averin and E.V.\ Sukhorukov,
 Phys.\ Rev.\ Lett.\ {\bf 95}, 126803 (2005).

\bibitem{muzykantskii:03}
B.A.\ Muzykantskii and Y.\ Adamov,
 Phys.\ Rev.\ B {\bf 68}, 155304 (2003).

\bibitem{shelankov:03}
A.\ Shelankov and J.\ Rammer,
 Europhys.\ Lett.\ {\bf 63}, 485 (2003).

\bibitem{avron:08}
J.E.\ Avron, S.\ Bachmann, G.M.\ Graf, and I.\ Klich,
 Commun.\ Math.\ Phys.\ {\bf 280}, 807 (2008).

\bibitem{roothaan:51}
C.C.J.\ Roothaan,
 Rev.\ Mod.\ Phys.\ {\bf 23}, 69 (1951).

\bibitem{familiar}
In condensed matter theory this statement is made use of in the
  Bogoliubov-Valatin transformation where the quadratic Hamiltonian is
  diagonalized while keeping the commutation relations invariant.

\bibitem{courant}
R.\ Courant and D.\ Hilbert,
 {\em Methoden der mathematischen Physik},
 4th edition (Springer-Verlag, Berlin, Heidelberg, New York, 1993).

\bibitem{lebedev:08}
M.V.\ Lebedev, A.A.\ Shchekin, and O.V.\ Misochko,
 Quantum Electron.\ {\bf 38}, 710 (2008).

\bibitem{abanov:08}
A.G.\ Abanov and D.A.\ Ivanov,
 Phys.\ Rev.\ Lett.\ {\bf 100}, 086602 (2008).

\bibitem{dejong:94}
M.J.M.\ de~Jong and C.W.J.\ Beenakker,
 Phys.\ Rev.\ B {\bf 49}, 16070 (1994).

\bibitem{bodoky:08}
F.\ Bodoky, W.\ Belzig, and C.\ Bruder,
 Phys.\ Rev.\ B {\bf 77}, 035302 (2008).

\bibitem{timeorder}
We assume that the Hamilton operators $\mathcal{H}(t)$ and $\mathcal{H}(t')$
  for different times $t\neq t'$ commute. Otherwise a time-ordering operator
  has to be introduced.

\bibitem{klich:03}
I.\ Klich,
 in Ref.\ \onlinecite{nazarov}.

\bibitem{muzykantskii:94}
B.A.\ Muzykantskii and D.E.\ Khmelnitskii,
 Phys.\ Rev.\ B {\bf 50}, 3982 (1994).

\bibitem{nazarov:02}
Yu.V.\ Nazarov and D.A.\ Bagrets,
 Phys.\ Rev.\ Lett.\ {\bf 88}, 196801 (2002).

\bibitem{pilgram:03}
S.\ Pilgram and M.\ B\"uttiker,
 Phys.\ Rev.\ B {\bf 67}, 235308 (2003).

\bibitem{brown:89}
R.J.\ Brown, M.J.\ Kelly, M.\ Pepper, H.\ Ahmed, D.G.\ Hasko, D.C.\ Peacock,
  J.E.F.\ Frost, D.A.\ Ritchie, and G.A.C.\ Jone,
 J.\ Phys.: Condens.\ Matter {\bf 1}, 6285 (1989).

\bibitem{basor:79}
E.L.\ Basor,
 Indiana Univ.\ Math.\ J.\ {\bf 28}, 975 (1979).

\bibitem{schottky:18}
W.\ Schottky,
 Ann.\ Phys.\ (Leipzig) {\bf 57}, 541 (1918).

\bibitem{simon:05}
B.\ Simon,
 in {\em Geometry, Spectral Theory, Groups, and Dynamics},
 vol.\ 387 of {\em Contemporary Mathematics}, pp.\ 253--275 (American
  Mathematical Society, Providence, 2005).

\end{thebibliography}
\end{document}